\documentclass[traditabstract]{aa}
\usepackage{txfonts}
\usepackage{amsmath}
\usepackage{color}
\usepackage{longtable}
\bibpunct{(}{)}{;}{a}{}{,} 

\newcommand{\kms}{km~s$^{-1}$\,}

\newcommand {\vi} {{\it V--I\/}}
\newcommand {\vh} {{\it V--H\/}}
\newcommand {\vj} {{\it V--J\/}}
\newcommand {\vk} {{\it V--K\/}}

\newcommand {\teff} {T$_{\rm{eff}}$\/}
\newcommand {\teffc} {T$_{\rm{eff,col}}$\/}
\begin{document}

\title{Carbon and nitrogen abundances of individual stars in the 
Sculptor dwarf spheroidal galaxy\thanks{Based on observations made with ESO Telescopes at the La Silla Paranal Observatory under programme ID 091.D-0089}}
\subtitle{} 

\author{C. Lardo\inst{1}, G. Battaglia\inst{2,3}, E. Pancino\inst{4,5}, D. Romano\inst{4}, T.~J.~L. de Boer\inst{6}, E. Starkenburg\inst{7}, E. Tolstoy\inst{8}, M.~J. Irwin\inst{6}, P. Jablonka\inst{9,10}, and M. Tosi\inst{4} }

\institute{Astrophysics Research Institute, Liverpool John Moores University, 146 Brownlow Hill, Liverpool L3 5RF, UK 
\email{C.Lardo@ljmu.ac.uk}
\and 
 Instituto de Astrofisica de Canarias, 38205 La Laguna, Tenerife, Spain
 \and
  Universidad de la Laguna, Dpto. Astrofisica, 38206 La Laguna, Tenerife, Spain 
  \and
  INAF Osservatorio Astronomico di Bologna, via Ranzani 1, 40127 Bologna, Italy 
  \and
  ASI Science Data Center, Via del Politecnico SNC, 00133, Roma, Italy
   \and
 Institute of Astronomy, University of Cambridge, Madingley Road, Cambridge CB3 0HA, UK
 \and
 Leibniz-Institute for Astrophysics Potsdam (AIP), An der Sternwarte 16, 14482 Potsdam, Germany
 \and
 Kapteyn Astronomical Institute, University of Groningen, PO Box 800, 9700AV Groningen, The Netherlands 
 \and
 Laboratoire d' astrophysique, Ecole Polytechnique F\'{e}d\'{e}rale de Lausanne (EPFL), Observatoire, CH-1290 Versoix, Switzerland
 \and
 GEPI, Observatoire de Paris, CNRS, Universit\'{e} de Paris Diderot, F-92195 Meudon, Cedex, France
 \\}
\date{Received / Accepted}

\abstract{
We present [C/Fe] and [N/Fe] abundance ratios and CH($\lambda$4300) and S($\lambda$3883)
index measurements for 94 red giant branch (RGB) stars in the Sculptor dwarf spheroidal galaxy
from VLT/VIMOS MOS observations at a resolving power R= 1150 at 4020~\AA.
This is the first time that [N/Fe] abundances 
are derived for a large number of stars in a dwarf spheroidal.
We found a trend for the [C/Fe] abundance to decrease with increasing luminosity
on the RGB across the whole metallicity range, a phenomenon observed in
both field and globular cluster giants, which can be interpreted in the framework of evolutionary mixing 
of partially processed CNO material. Both our measurements of [C/Fe] and [N/Fe] are
in good agreement with the theoretical predictions for stars
at similar luminosity and metallicity. We detected a dispersion in the carbon abundance at a given [Fe/H], which cannot be ascribed to 
measurement uncertainties alone. We interpret this observational evidence
as the result of the contribution of different nucleosynthesis sources over time to a not well-mixed interstellar medium.
We report the discovery of two new carbon-enhanced, metal-poor stars. These are
likely the result of pollution from material enriched by asymptotic giant branch stars,
as indicated by our estimates of [Ba/Fe]$> +1$.
We also attempted a search for dissolved globular clusters in the field
of the galaxy by looking for the distinctive C-N pattern of second population
globular clusters stars in a previously detected, very metal-poor,
chemodynamical substructure. We do not detect chemical
anomalies among this group of stars. However,  small number statistics and limited spatial
coverage do not allow us to exclude the hypotheses that this substructure forms part
of a tidally shredded globular cluster. }


\titlerunning{C and N abundances in Sculptor dSph}
\authorrunning{Lardo et al.}
\maketitle

\section{Introduction}
Carbon and  nitrogen are among the most abundant metals in the Universe 
and they have a fundamental role in the chemical evolution of galaxies. 
Carbon is produced in stars of all masses, essentially by helium burning. 
Nitrogen is synthesised during hydrogen burning through the CNO cycle. 
While some N production is predicted in massive stars by
mixing between the helium-burning 
and the hydrogen-burning layers, a large amount of N 
is made in intermediate-mass stars of 
4 to 11~M$_{\sun}$, which undergo hot-bottom burning \citep{renzini81,siess10}.
The abundances of both C and N are changed
significantly as the stars evolve along the 
red giant branch (RGB) because of internal mixing, which 
brings material processed through the CNO cycle in the stellar interiors
to the photosphere. As a result, in evolved RGB stars, C is depleted and N is enhanced relative
to main- sequence or subgiant branch stars that had the same chemical composition at birth.
Therefore, the interpretation of the observations in terms of evolution of both elements needs to account for these 
stellar evolution effects.

While the abundance properties of C and N in individual stars in 
globular clusters (GCs) have been widely explored, little is known of these properties in 
galactic systems of not too dissimilar a stellar mass, i.e., the early-type dwarf galaxies that 
surround the Milky Way (MW). Given the abundance of dwarf galaxies in the Local Group,
which is conveniently placed at a relatively close distance to us,
using current facilities the stellar populations of these galaxies can be studied on a star-by-star
basis. They are also 
the galactic environments with the lowest mean metallicity.
Thus, the study of their stellar populations gives us the possibility to learn about star
formation processes in the most pristine environments we can observe \citep{tolstoy09}, acting as the local, accessible counterparts of high-redshift objects.

Sculptor is one of the most studied dwarf spheroidal (dSph) galaxies.
It is relatively faint (M$_{\rm{V}}$ $\simeq$ --11.2) and
located at a distance of 86 $\pm$ 5 kpc \citep{pietr08}.
Star formation in this galaxy took place 
for an extended period of time, although it
mostly contains stars older than 10 Gyr \citep{deboer12}. This
makes Sculptor an ideal candidate to study the earliest phases of star formation and chemical
enrichment processes.
As such, a number of large spectroscopic surveys have been carried out in this galaxy. 
Metallicities ([Fe/H]) from intermediate resolution spectroscopy 
of several hundred RGB stars, which are probable members of Sculptor, were determined by
\citet{tolstoy04} and \citet{battaglia08b} out to its
nominal tidal radius. An extended sample of Scl stars is presented in
\citet{starkenburg10}. \citet{kirby09} derived abundances
for Fe, Mg, Si, Ca, and Ti from medium-resolution  (R$\simeq$6500) spectroscopy
for 388 radial velocity member stars within $\sim$2 core radii.  
Abundances have been measured for a range of elements, including Fe, Ca, Mg, Ti, Y, Ba and Eu for about 90 stars in Sculptor from high-resolution FLAMES/GIRAFFE observations in the central regions (\citealp{tolstoy09}; Hill et al. in prep.)

Additionally, carbon has been derived in many giants in this dSph.
\citet{azzopardi85,azzopardi86} survey Sculptor for carbon- and CH-stars, but they did not quantify their carbon enhancement (see also 
\citealp{groenewegen09}). \citet{shetrone98} measured low-resolution [C/Fe] abundances for two stars in Sculptor, confirming that the CN strong star found by \citet{smith83} exhibits both strong CN and CH bands compared to other stars with similar atmospheric parameters, but does not show strong C$_{2}$ absorptions. They conclude that this star is not similar to the CN strong stars found in GCs, where 
CN strong stars usually show weak CH absorption.
High-resolution spectroscopic follow up of the most metal-poor candidate stars in Sculptor
was carried out by \citet{tafelmeyer10}, \citet{frebel10}, \citet{kirby12}, \citet{starkenburg13},  and \citet{simon15};\ these studies  
 include C measurements. No carbon-enhanced metal poor (CEMP) stars were identified in those studies.
Recently, \citet{skuladottir15} estimated carbon abundances and upper limits
from near-infrared CN bands for about 80 Sculptor giants observed with FLAMES/GIRAFFE. They also 
discovered the first CEMP-no star in Scl, i.e., with no overabundance of n-capture elements.
Finally, \citet{kirby15} present [C/Fe] abundances for
several globular clusters and dSph RGB stars, including 198 stars in Sculptor.
They used these abundances to study evolutionary effects on carbon abundances in two 
different environments (i.e., dSph and GCs), Sculptor chemical enrichment, and its 
relation with the Galactic halo.

In this work, we report carbon and nitrogen abundances of 94 RGB stars in Sculptor. This samples adds 
to the literature C- measurements for 85 new stars; 
to our knowledge, none of the large spectroscopic surveys of Sculptor have so far included measurements
of nitrogen abundances, with the exception of one star in \citet{skuladottir15}. 

We also use our VIMOS data to carry out the first attempt for a search of dissolved globular clusters
in the field of a dwarf spheroidal galaxy via the investigation of chemical anomalies.
Only a handful of Local Group dwarf galaxies host GCs at present \citep{vandenbergh06}; it is however conceivable that a 
larger fraction of Local Group dwarf galaxies may have once hosted GCs, now in the form of disrupted remnants.
 This is suggested by the presence
 of cold-kinematic {substructures} in several of these objects,
 i.e. groups of stars with a very low velocity dispersion with respect to 
the overall population of the galaxy 
\citep[e.g., in Ursa Minor, Sextans and Sculptor dSphs; ][]{kleyna03,battaglia07,battaglia11}. 
Interestingly, the metallicity measurements existing for individual stars in
Sculptor and Sextans suggest very low average metallicities ([Fe/H]=--2.0, --2.5 dex) for these substructures,
which would place them among the most metal-poor stellar clusters known. 

The presence of disrupted GCs in dwarf galaxies provides insight into the star formation modes at
low metallicity/early times. In the Fornax dSph, a comparison between the metallicity distribution function of
field stars and that of the surviving GCs  already revealed that one-fifth to one-fourth of all stars in the Fornax dSph with
[Fe/H] $< -2$ dex belong to the four most metal-poor GCs. This also implies that these GCs could not have been four to five times
more massive than at present, which poses difficulties for scenarios 
of self-enrichment and early evolution of GCs postulating much larger masses for their progenitors \citep{larsen12}. 

\begin{figure*}
\begin{center}
\includegraphics[width=12.5cm, angle=90]{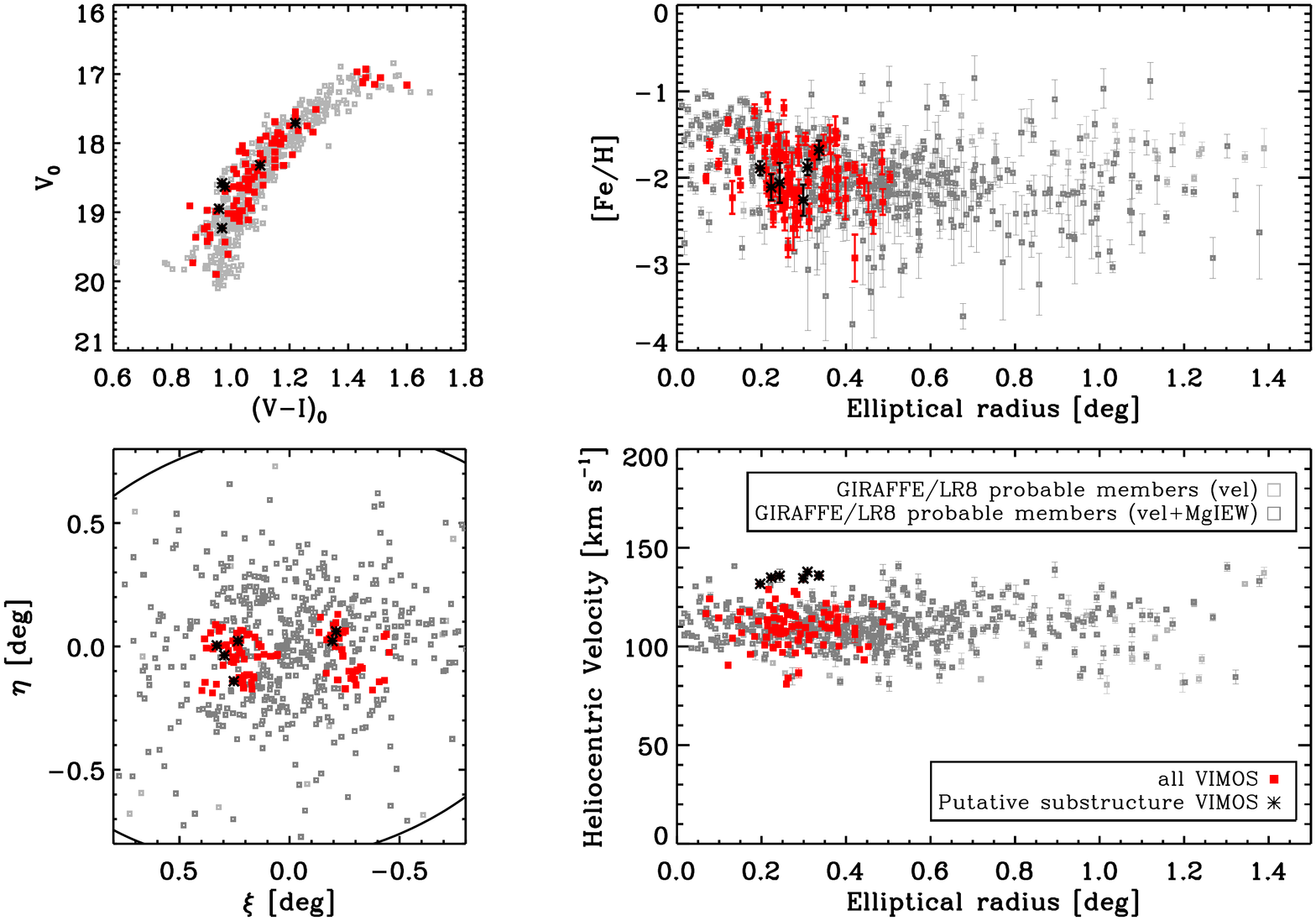}
\caption{Properties of our VIMOS sample of highly, likely member stars to Sculptor compared to 
those of the DART data set of Sculptor stars (see legend). From the top left panel in anticlockwise direction, the location of our 
VIMOS spectroscopic targets is shown on the Sculptor color-magnitude diagram, field-of-view,  l.o.s. velocity versus projected radius and metallicity  
versus projected radius planes.}
\label{fig:CMD}
\end{center}
\end{figure*}

While chemical tagging is a promising route for identifying stars that share very similar chemical 
signatures (e.g., \citealp{karlsson12} for an application to Sextans), {the so-called unambiguous 
signature} that a star formed in a GC-like environment is the presence of 
characteristic anticorrelations between light elements such as C, N, O, and Na 
(see, e.g., \citealp{smith84}; \citealp{carretta09}; \citealp{martell09}; \citealp{martell10}; and  \citealp{gratton12} for a recent review).
These kinds of anticorrelations are currently known to be present in GCs with present-day mass as
low as ~10$^{4}$ M\sun~(e.g., \citealp{dalessandro14}), of the order of the stellar mass estimated for the
substructures in Sextans and Sculptor.
Here we focus on the chemodynamical substructure detected by \citet{battaglia07} in Sculptor as a group of stars clustering in a 
narrow velocity range well distinct from Sculptor systemic velocity at projected elliptical radii
0.18$\lesssim$ R[deg] $\lesssim$0.35. We revisit this classification by assigning a probability of 
membership to the substructure according to the line-of-sight  velocity distribution (LOSVD) expected for MW,
Sculptor field, and 
Sculptor substructure stars at 0.18 $\le$ R[deg] $\le$0.35. We model the overall LOSVD as a sum of Gaussians 
 along the lines of studies by 
\citet{battaglia07}, but add the substructure as am additional term. This yields an estimated central velocity and
dispersion of the Sculptor main body and substructure $v_{\rm main} = 109.8 $\kms, $\sigma_{\rm main} = 7.4 $\kms,
and $v_{\rm sub} = 133.9 $\kms and $\sigma_{\rm sub} = 3.3 $\kms, respectively. 
Seven stars are identified as probable substructure members: they have  
an average CaT [Fe/H] $= -1.9 \pm 0.1$ dex, with a scatter of 0.17$\pm$0.1 dex, 
which is very close to the average [Fe/H] error of 0.14 dex.
We were able to allocate VIMOS slits on six of these substructure stars, and their 
location on the CMD, field-of-view, metallicity versus R and line-of-sight (l.o.s.) velocity versus R
is plotted in Fig.~\ref{fig:CMD}.

The faintness of the substructure stars (down to V$\sim$19.3) makes it very hard to search for Na-O anticorrelations 
because determining the oxygen abundance from, for example the small oxygen lines around 6300 or 7700~\AA,\,
would require prohibitive exposure times even on 8-m class telescopes.
Next, we  focus on the search for C and N anticorrelations,
following \citet{martell10} and \citet{martell11} for SEGUE spectra of Milky Way halo stars. 
To disentangle chemical anomalies due to the stars being born in a GC environment from the modification to the 
C and N abundances induced by mixing in luminous giant stars, it is  crucial to 
also establish the trend of C and N abundances for the field population.
Then, we  build a suitable comparison
sample among our 94 RGB stars by selecting Sculptor field RGB stars spanning a similar range in metallicity,
effective temperature, and luminosity to those in 
the substructure.

This paper is organised as follows.
Section~\ref{OBSERVATIONS} presents the observational material and data reduction. 
We outline our analysis procedure in Sect.~\ref{ANALYSIS}.
In Sect.~\ref{RESULTS} we discuss our results on the C- and N- abundance determinations.
We summarise our results and present conclusions in Sect.~\ref{SUMMARY}. 

\section{Observations and data reduction}\label{OBSERVATIONS}

The coordinates, V and I magnitudes, l.o.s. velocities, and [Fe/H]
of our sample of 94 stars are known
from the DART VLT/FLAMES GIRAFFE spectroscopic survey of Sculptor in the region of the NIR Ca~II triplet lines 
\citep{tolstoy04,battaglia08b} and from CTIO/MOSAIC~II photometric observations \citep{deboer11}. 

Ninety-two targets are highly likely to be RGB stars members of Sculptor, 
as  indicated by both their l.o.s. velocities and Mg~I equivalent width  
(see \citealt{battaglia12}). For the remaining two stars 
(flagged in Table~\ref{tab:tabella1}), the Mg~I equivalent width is just outside the 
range considered by \citet{battaglia12} for RGB stars;
we keep these two objects in the analysis as they do not display properties 
discrepant from the rest of the sample. 

We cannot exclude that our sample contains some contamination from asymptotic giant branch (AGB) stars, as the AGB evolutionary stage merges with the RGB in the CMD. However, it is most likely that all stars in Scl are older than 10 Gyr \citep{deboer12}, thus we do not expect severe contamination by intermediate-age AGB stars.

The observations were carried out with VLT/VIMOS in July 2013 in visitor mode, 
with relatively good atmospheric conditions, e.g., the typical seeing was around 1.0-1.2\arcsec. 
For these MOS observations, we used the HR-blue-(NEW) grism 
with a slit width of 1.2\arcsec, yielding a resolving power R $\simeq$ 1100 at 4020~\AA. The VIMOS field-of-view consists 
of four quadrants of 7\arcmin~$\times$~8\arcmin each, separated by a gap of about 2\arcmin. For an object at the 
centre of the quadrant, the wavelength range covered by the HR-blue-(NEW) grism ranges from 3700 to 5240~\AA. 
When planning the observations, we paid 
particular attention to including the CN-band at 3800~\AA, in the probed wavelength range of each target. 

The location of our VIMOS sample on the Sculptor 
color-magnitude diagram and field-of-view is shown in Fig.~\ref{fig:CMD}, 
where  its metallicity and l.o.s. velocity properties 
are also compared to the overall population of Sculptor RGB probable members. 

We observed four MOS masks for a total integration time of 18.3 hours; the signal-to-noise ratio (SNR) per pixel 
ranges from 5 to 50 in the CN feature 
region (at 3880~\AA; see Table~\ref{tab:tabella1}). 
Seven stars were observed in more than one mask to test the internal measurement errors. 
The data reduction was performed with the ESO pipeline\footnote{\url{http://www.eso.org/sci/software/pipelines/}}
with the standard settings described in the pipeline manual. The pipeline processing steps included the
subtraction of the median-combined bias image, creation of the
spectral extraction mask from a flat-field image taken immediately
following the science exposure, and wavelength calibration
constructed from the HeArNe lamp exposures.

We used the radial velocity as derived by the DART VLT/FLAMES GIRAFFE spectroscopic survey 
\citep{tolstoy04} to correct for radial velocity shifts. We did not attempt to detect possible 
binary systems by measuring radial velocity from the VIMOS spectra because
at this relatively low resolution, the internal error estimated from a  cross-correlation against a template, 
as described in \citet{lardo12}, is larger ($\simeq$30-40 km s$^{-1}$) than radial velocity variations expected 
in binary stars. The remaining small shifts are furthermore amended by performing 
a cross-correlation of the object spectrum with a synthetic spectrum using the most prominent spectral features 
(e.g., H$\beta$, H$\gamma$, H$\delta$, and Ca H+K) in our wavelength range.

\begin{figure}
\begin{center}
\includegraphics[width=0.75\columnwidth]{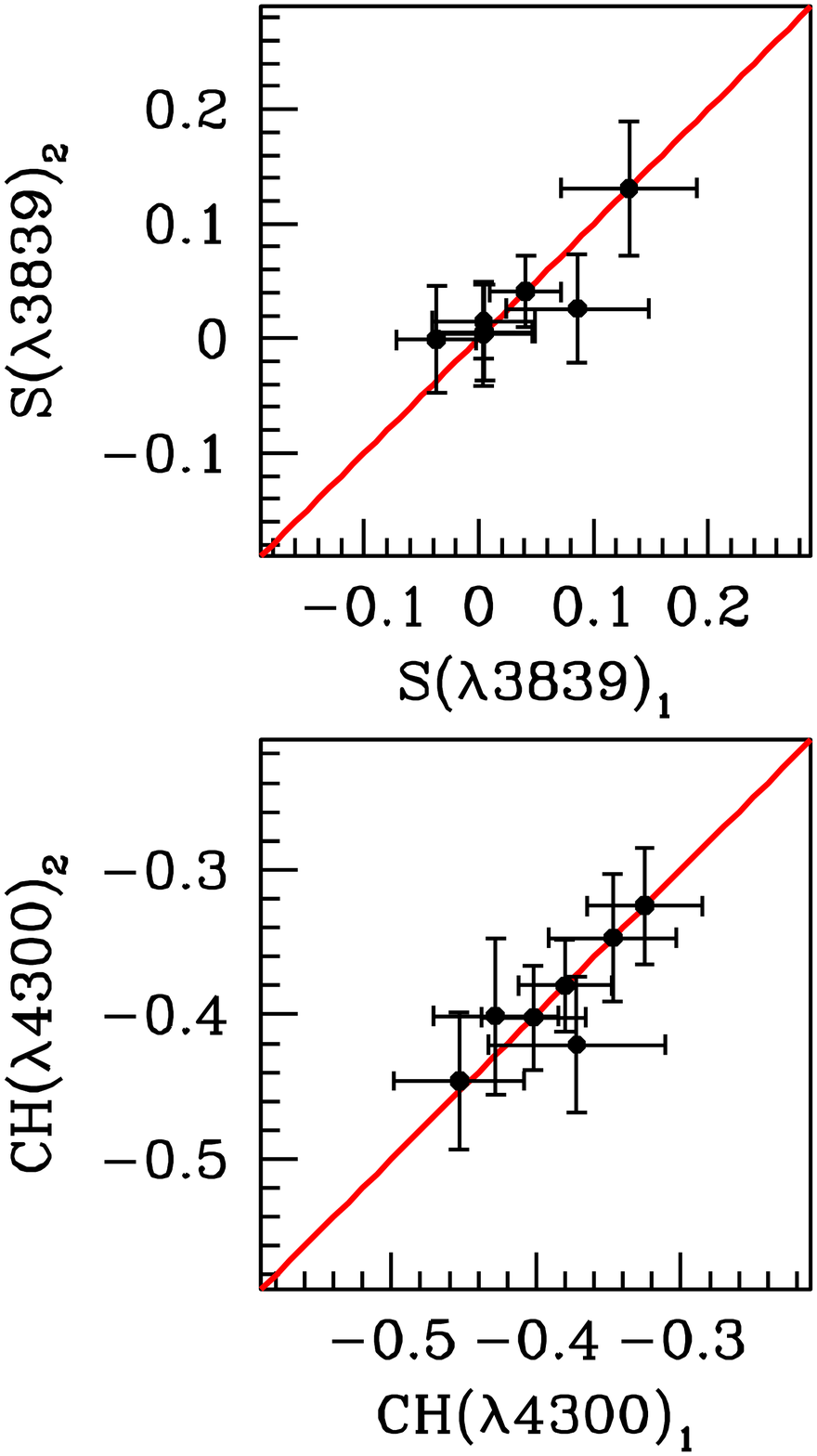}
\caption{Comparison of our CN and CH measured indices for stars with multiple observations. 
The red line shows the 1:1 relation. }
\label{fig:CHECK}
\end{center}
\end{figure}

\begin{figure*}[htbp]
\begin{center}
\includegraphics[width=15cm]{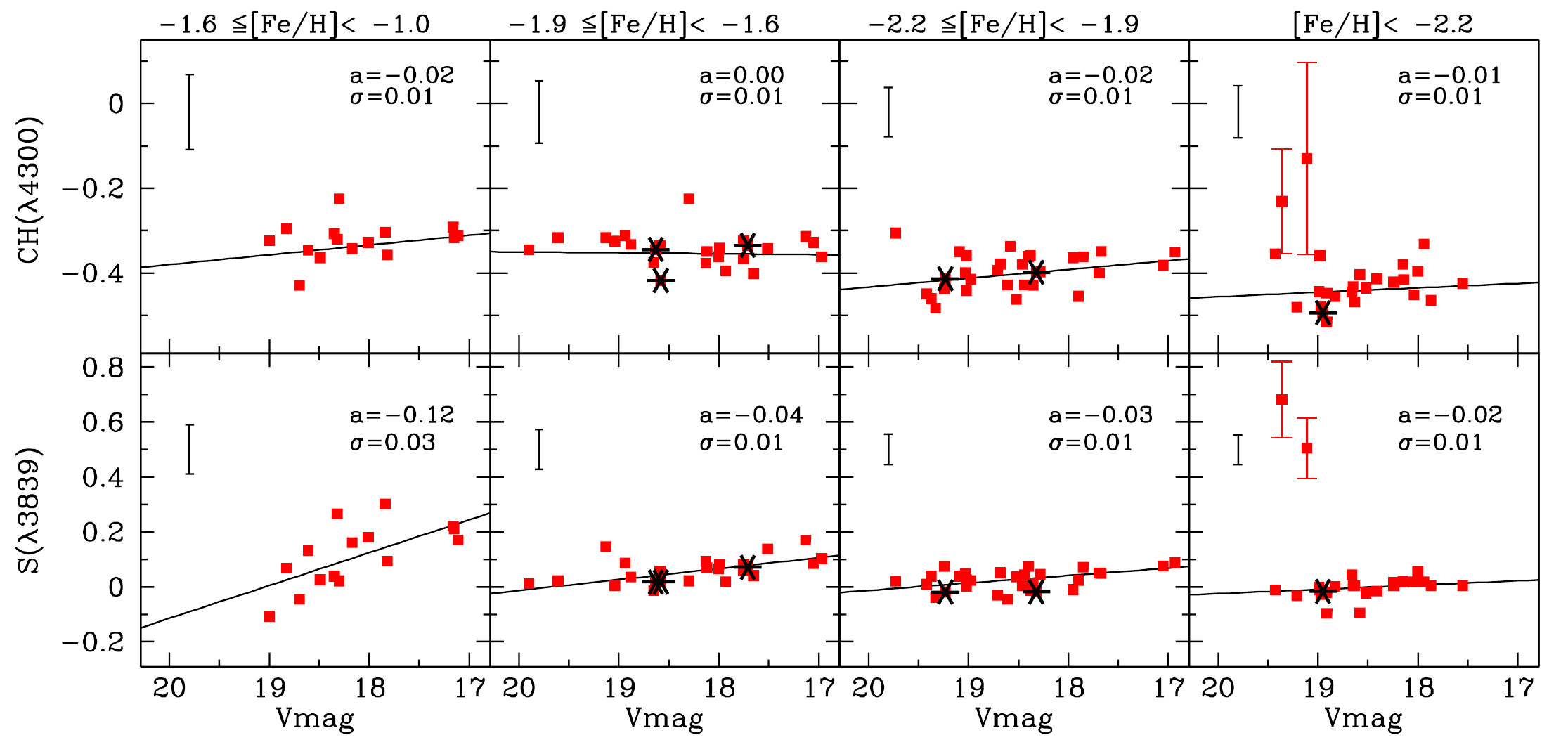}
\caption{CH($\lambda$4300) and S($\lambda$3839) band-strength index versus extinction-corrected apparent visual magnitude for 
probable member Sculptor RGB stars in four metallicity bins. Black asterisks denote objects belonging 
to the kinematic  substructure (see text). The range of metallicities considered for each bin is given in the upper part of each panel. 
We report the mean error associated with index measurements in the top left corner of each panel. 
Solid lines are best fits to the data when the two notable S($\lambda$3839) outliers in
the most metal-poor bin are excluded. The slope of the best-fit line (a) and its associated error ($\sigma$) are also reported in the upper right corner of each panel. Error bars for the stars with strong CN absorption are explicitly given in the last panel.}

\label{fig:INDEX}
\end{center}
\end{figure*}

\section{Indices and abundance analysis}\label{ANALYSIS}

\subsection{Spectral indices}\label{IND}
In a first approximation,  the CH($\lambda$4300) index can be considered as a carbon-sensitive diagnostic, 
while the S($\lambda$3839) index follows N abundance. 
In a star with a solar C/N ratio (4.2/1) 
the nitrogen abundance  drives the formation of the CN band
because the molecular abundance is governed by the abundance of the minority species. 
However, C and N abundances are continuously changed by mixing 
during RGB evolution.
As a result, the C/N ratio declines while 
the CN feature becomes stronger because of the increase of the $^{14}$N abundance. 
The maximum in the CN strength is reached when the C/N ratio approaches unity. 
Once the C/N ratio decreases below unity, carbon becomes the minority species controlling the CN band
formation so that any additional drop in the C abundance results in weaker CN bands.
In this section, we measure a set of indices quantifying the strength of the UV/CN band, S($\lambda$3839) and the G band 
of CH, CH($\lambda$4300). 

We adopt the same indices as defined in \citet{pancino10} and \citet{lardo12,lardoM2,lardo13}; the uncertainties related to the index 
measurements are obtained with the expression derived by \citet{Vollmann06}, assuming pure photon 
noise statistics in the flux measurements. The indices, together with additional information on the target stars, 
are listed in Table~\ref{tab:tabella1}.
As there are seven stars observed in different exposures, 
we compare the CH($\lambda$4300) and S($\lambda$3839) indices obtained from different spectra of the same star. 
As demonstrated in Fig.~\ref{fig:CHECK}, the agreement between the index measurements is excellent. 
This validates both the data reduction and the spectral indices analysis.

 The strength of the S($\lambda$3839) index strongly depends on
the overall metallicity of the star because CN is a double-metal molecule. Spectroscopic targets in Sculptor span a range of [Fe/H]$\simeq$1.8 dex in metallicity. Therefore, 
we must isolate 
C and N abundance variations from metallicity effects. To this end, we arbitrarily define four metallicity groups: 
very metal-poor  ([Fe/H] $<$ --2.2 dex), 
metal-poor (--2.2 $\leq$ [Fe/H] $<$ --1.9 dex), intermediate metallicity (--1.9$\leq$ [Fe/H] < --1.6 dex), and metal-rich stars (--1.6 $\leq$ [Fe/H] < --1.0 dex).
 The S($\lambda$3839) and CH($\lambda$4300) index measurements
in the four metallicity bins are plotted in Fig.~\ref{fig:INDEX} against the visual magnitude.

\begin{figure}
\begin{center}
\includegraphics[width=\columnwidth]{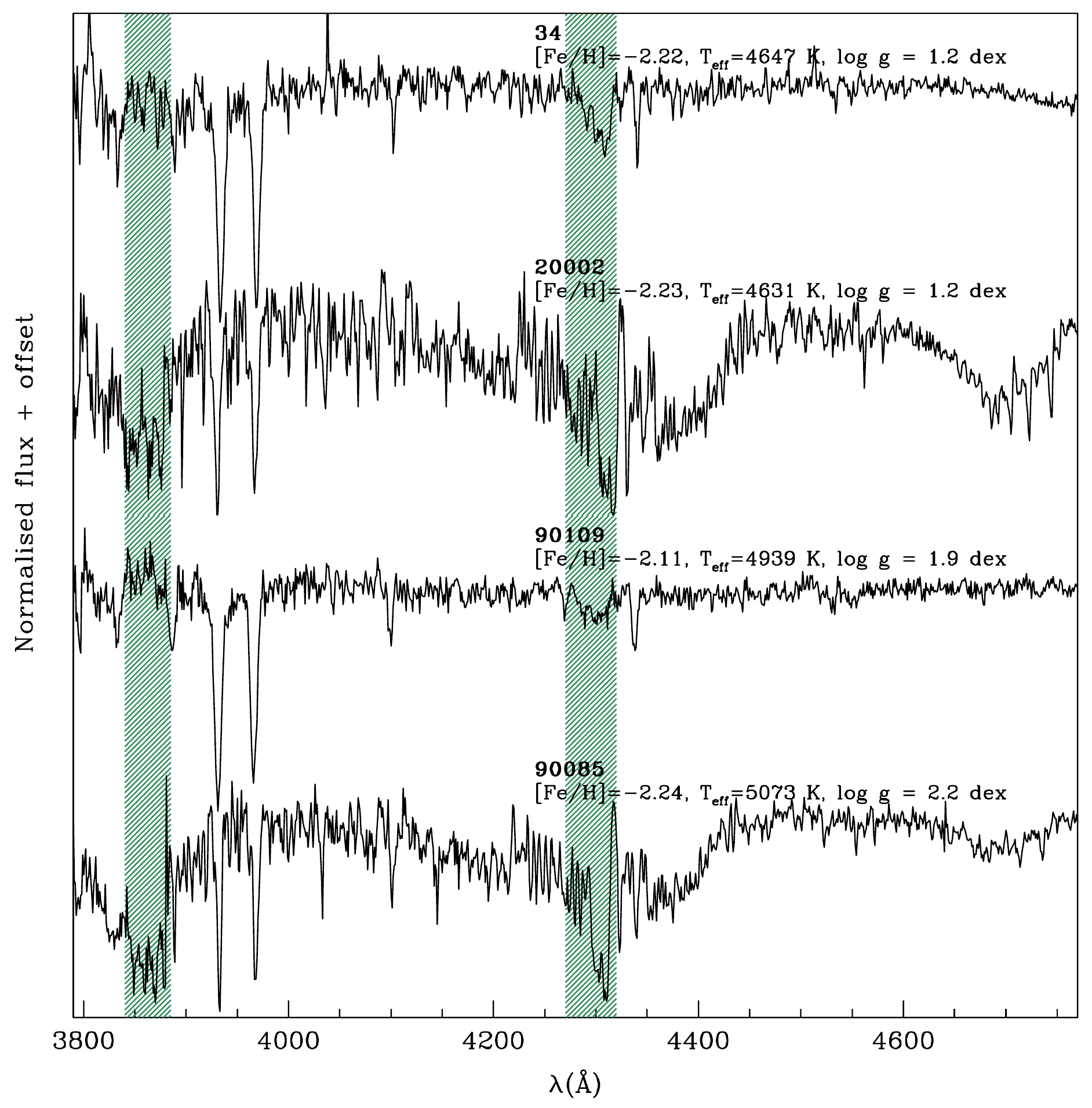}
\caption{Comparison between the spectra of the stars with large C abundance, 
stars 20002 and 90085, and two sample stars with similar parameters yet very different CN and CH absorptions.
Also,  the narrow Ca(H+K) absorption lines indicate very low metallicity ([Fe/H] $\simeq$ -2.2 dex).
The green hatched regions are the windows from which we measured the CN and CH indices. The strong features at 
4380\AA~and 4738\AA~are C$_{2}$ Swan absorptions. }
\label{fig:SPETTRI}
\end{center}
\end{figure}

Two stars have prominent S($\lambda$3839) absorptions in the most metal-poor bin: star 20002 and 90085 
with S($\lambda$3839)= 0.50 $\pm$ 0.11 and 0.68 $\pm$ 0.14 mag, respectively. These stars also have a large CH($\lambda$4300) index, 
CH($\lambda$4300)= --0.13 $\pm$ 0.23 and --0.23 $\pm$ 0.12 mag, although consistent within 1-$\sigma$ with the rest 
of the Sculptor population in the same metallicity bin. The spectra of the two CN-strong stars are  
compared to stars of similar stellar atmosphere parameters in Fig.~\ref{fig:SPETTRI}.
It is clear that the enhanced CH and CN index is not an artefact due to,  
for example, a low SNR. In the figure,  the data quality and  some of the spectral features of 
interest for the present analysis are also apparent. In Sect.~\ref{RESULTS} we discuss these two objects
in more details. 

To explore the presence of trends with magnitudes, we perform weighted fits to the data 
in the index-magnitude plane, when excluding the two outlier stars with high S($\lambda$3839) index 
in the most metal-poor bin (bottom right panel).
No statistically significant trend of CH($\lambda$4300) index with $V$ magnitude is detected (see slopes and errors as 
given in Fig.~\ref{fig:INDEX}); while we find a trend of increasing S($\lambda$3839) index for brighter stars, 
with the relation being steeper for the highest metallicity bin and flattening at lower metallicity.

The pattern of C and N abundances underlying the CH and CN band indices of Fig.~\ref{fig:INDEX}
cannot be interpreted on the basis of band strengths alone because both indices 
are sensitive to atmospheric parameters and metallicity of the stars. 
Since we are using the CN band,  the coupling between 
the C and N abundances must also be taken into account. 
In the following sections, we perform spectral synthesis to disentangle the underlying
C and N abundances from the CH and CN band strengths.

\begin{longtab}
\setlength{\tabcolsep}{0.11cm}
\begin{longtable}{lccrrrrcrrccc}
\caption{Sample of probable Sculptor members with properties and index measures. We label the stars as ``sub'' if they were considered as part of the metallicity-substructure. Notes: 1= MgIEW just beyond the range for Sculptor giants considered in \citet{battaglia12}; 
2= photometric flag suggests nature as extended object; visual inspection of DSS and 2mass images suggests 147 
is a blend, while 90109 looks just as a point-source. \label{tab:tabella1}}\\
\hline \hline
ID             &     RA   &    DEC      &         V       &  [Fe/H]     &CN     &  eCN    & SNR$_{\rm{CN}}$ & CH    & eCH   &  SNR$_{\rm{CH}}$  & label  & Notes\\
                 &   (J2000)   &   (J2000)  &   (mag)  &  (dex)       &(mag)  & (mag)   &                            & (mag) & (mag) &                                &               &        \\ 
 \hline
\endfirsthead
\caption{continued.}\\

\hline
\hline
ID             &     RA   &    DEC      &         V       &  [Fe/H]     &CN     &  eCN    & SNR$_{\rm{CN}}$ & CH    & eCH   &  SNR$_{\rm{CH}}$  & label  & Notes\\
                 &   (J2000)   &   (J2000)  &   (mag)  &  (dex)       &(mag)  & (mag)   &                            & (mag) & (mag) &                                &               &        \\ 
\hline
\endhead
\hline
\endfoot
   11     & 00:58:58.88 & --33:39:02.5 & 18.15 & --2.25&  0.021  & 0.055 & 14  &    --0.38   & 0.055 & 26  & &       \\   
  12     & 00:58:56.16 & --33:38:42.1 & 18.68 & --2.19&  0.052  & 0.074 & 10  &    --0.379  & 0.065 & 22  & &       \\
  15     & 00:59:08.60 & --33:42:29.4 & 17.95 & --2.12&  --0.01  & 0.08  & 10  &    --0.364  & 0.072 & 20  & &     \\
  17     & 00:59:04.05 & --33:40:31.4 & 18.0  & --2.42&  0.056  & 0.078 & 17  &    --0.3965 & 0.077 & 30  & &       \\
  19     & 00:59:00.27 & --33:43:14.4 & 17.94 & --2.27&  0.018  & 0.079 & 10  &    --0.331  & 0.08  & 18  & &       \\
  20     & 00:59:06.12 & --33:50:34.2 & 18.24 & --2.39&  0.0040 & 0.03  & 26  &    --0.422  & 0.029 & 48  & &       \\
  21     & 00:58:39.96 & --33:48:09.8 & 18.58 & --2.27&  --0.094 & 0.056 & 14  &    --0.404  & 0.053 & 26  & &     \\
  22     & 00:58:19.97 & --33:53:02.4 & 18.83 & --2.28&  0.0020 & 0.04  & 19  &    --0.456  & 0.033 & 43  & &       \\
  25     & 00:58:05.37 & --33:39:57.2 & 18.58 & --2.04&  0.123  & 0.076 & 10  &    --0.337  & 0.069 & 21  & &       \\   
  28     & 01:01:29.21 & --33:43:30.1 & 17.55 & --2.44&  0.0050 & 0.031 & 25  &    --0.425  & 0.03  & 47  & &       \\   
  30     & 01:01:17.80 & --33:39:14.8 & 18.61 & --1.97&  --0.045 & 0.051 & 15  &    --0.428  & 0.079 & 18  & &     \\
  31     & 01:01:31.76 & --33:42:07.1 & 18.91 & --2.41&  --0.021 & 0.034 & 22  &    --0.448  & 0.041 & 34  & &     \\
  32     & 01:01:23.15 & --33:41:05.1 & 18.38 & --2.04&  --0.014 & 0.061 & 13  &    --0.359  & 0.073 & 19  & &     \\
  34     & 01:01:36.83 & --33:37:46.6 & 18.52 & --2.22&  --0.023 & 0.043 & 18  &    --0.436  & 0.053 & 26  & &     \\
  38     & 01:01:52.59 & --33:39:08.7 & 18.14 & --2.32&  0.016  & 0.045 & 17  &    --0.416  & 0.054 & 26  & &       \\
  40     & 01:01:48.10 & --33:42:30.6 & 18.97 & --2.24&  --0.028 & 0.048 & 16  &    --0.48   & 0.045 & 31  & &     \\
  42     & 01:01:49.33 & --33:43:02.9 & 18.44 & --2.09&  0.044  & 0.056 & 14  &    --0.428  & 0.049 & 29  & &       \\ 
  44     & 01:01:57.77 & --33:42:51.0 & 18.65 & --1.9 &  --0.014 & 0.052 & 15  &  --0.376  & 0.053 & 27  & &       \\ 
  46     & 01:02:00.00 & --33:51:13.3 & 17.75 & --1.88&  0.079  & 0.076 & 10  &    --0.324  & 0.074 & 19  & &       \\ 
  47     & 01:01:23.59 & --33:44:58.1 & 18.98 & --2.05&  0.023  & 0.03  & 26  &    --0.415  & 0.031 & 45  & & 2     \\
  49     & 01:01:25.35 & --33:46:09.6 & 19.03 & --1.96&  0.047  & 0.042 & 19  &    --0.399  & 0.042 & 33  & &       \\
  53     & 01:01:23.19 & --33:50:51.8 & 18.35 & --2.05&  0.012  & 0.065 & 12  &    --0.429  & 0.057 & 24  & &       \\
  56     & 01:01:14.17 & --33:43:59.1 & 19.02 & --2.11&  0.027  & 0.037 & 21  &    --0.36   & 0.035 & 41  & &       \\
  57     & 01:01:11.54 & --33:44:10.0 & 18.28 & --2.08&  0.045  & 0.03  & 25  &    --0.397  & 0.023 & 60  & &       \\
  60     & 01:01:13.18 & --33:38:34.5 & 18.41 & --2.49&  --0.015 & 0.055 & 14  &    --0.414  & 0.059 & 24  & &     \\
  61     & 01:01:10.28 & --33:38:37.9 & 17.85 & --2.16&  0.072  & 0.065 & 12  &    --0.362  & 0.08  & 18  & &       \\
  107    & 01:02:04.17 & --33:53:06.6 & 18.98 & --2.23&  --0.011 & 0.064 & 12  &    --0.36   & 0.049 & 29  & &     \\
  141    & 00:59:28.42 & --33:35:23.2 & 18.71 & --1.91&  --0.03  & 0.088 & 9   &    --0.394  & 0.078 & 18  & &     \\
  155    & 01:01:13.63 & --33:50:31.3 & 18.63 & --2.5 &  0.0030 & 0.043 & 18  &    --0.468  & 0.054 & 26  & &       \\
  156    & 01:01:08.14 & --33:49:32.2 & 19.21 & --2.32&  --0.032 & 0.095 & 8   &    --0.481  & 0.076 & 18  & &     \\
  158    & 01:00:56.41 & --33:49:47.1 & 18.65 & --2.46&  0.0030 & 0.015 & 51  &    --0.433  & 0.063 & 22  & &       \\ 
  182    & 00:58:48.66 & --33:50:16.0 & 18.46 & --1.97&  0.0040 & 0.064 & 17  &    --0.38   & 0.045 & 45  & & 1     \\ 
  183    & 00:58:55.98 & --33:52:46.0 & 19.02 & --2.02&  0.0010 & 0.043 & 18  &    --0.442  & 0.035 & 40  & &       \\   
  197    & 00:58:11.51 & --33:51:04.0 & 18.59 & --1.82&  0.056  & 0.042 & 19  &    --0.336  & 0.05  & 29  & &       \\
  198    & 00:58:04.07 & --33:50:41.7 & 17.67 & --1.99&  0.048  & 0.042 & 18  &    --0.349  & 0.044 & 32  & &       \\
  206    & 01:00:46.21 & --33:42:33.9 & 18.04 & --2.23&  0.018  & 0.052 & 15  &    --0.452  & 0.041 & 34  & &       \\
  208    & 01:00:33.86 & --33:44:54.4 & 17.75 & --1.84&  0.06   & 0.053 & 15  &    --0.367  & 0.059 & 24  & &       \\
  210    & 01:00:24.63 & --33:44:28.9 & 17.69 & --2.01&  0.05   & 0.048 & 16  &    --0.401  & 0.057 & 25  & &       \\
  220    & 00:59:30.49 & --33:39:04.0 & 17.9  & --2.09&  0.024  & 0.048 & 16  &    --0.455  & 0.041 & 34  & &       \\
  10001  & 00:59:07.84 & --33:38:48.0 & 18.32 & --2.11&  --0.018 & 0.068 & 11  &    --0.399  & 0.059 & 24  & sub&    \\ 
  10002  & 00:59:12.71 & --33:41:09.2 & 17.71 & --1.89&  0.071  & 0.065 & 12  &    --0.335  & 0.058 & 25  & sub&    \\ 
  10003  & 01:01:34.56 & --33:44:37.9 & 18.95 & --2.26&  --0.016 & 0.035 & 22  &    --0.494  & 0.031 & 45  & sub&    \\
  10004  & 01:01:16.50 & --33:41:09.8 & 19.23 & --2.06&  --0.019 & 0.058 & 21  &    --0.414  & 0.069 & 33  & sub&    \\ 
  20005  & 01:01:05.21 & --33:39:40.2 & 18.52 & --2.19&  0.037  & 0.045 & 17  &    --0.463  & 0.053 & 26  & &       \\   
  20001  & 01:01:01.44 & --33:49:20.4 & 18.4  & --2.14&  0.074  & 0.052 & 15  &    --0.361  & 0.057 & 25  & &       \\
  20002  & 01:01:15.03 & --33:50:02.6 & 19.11 & --2.23&  0.504  & 0.111 & 8   &    --0.13   & 0.227 &  7  & &       \\   
  20003  & 01:01:10.35 & --33:51:15.3 & 19.43 & --2.49&  --0.011 & 0.049 & 16  &    --0.355  & 0.071 & 20  & &     \\
  30001  & 01:01:44.23 & --33:42:12.3 & 18.58 & --1.68&  0.018  & 0.06  & 13  &    --0.418  & 0.059 & 24  & sub&    \\ 
  30002  & 01:01:22.15 & --33:50:55.0 & 18.63 & --1.88&  0.018  & 0.081 & 9   &    --0.345  & 0.088 & 16  & sub&    \\ 
  50001  & 00:58:59.03 & --33:44:47.4 & 18.3  & --1.6 &  0.022  & 0.187 & 4   &    --0.225  & 0.143 & 10  & &       \\
  50002  & 01:01:22.23 & --33:46:21.9 & 17.87 & --2.81&  0.0030 & 0.017 & 44  &    --0.465  & 0.022 & 64  & &       \\
  50004  & 01:00:58.23 & --33:52:51.6 & 19.09 & --2.15&  0.039  & 0.048 & 16  &    --0.35   & 0.057 & 25  & &       \\
  90015  & 00:59:11.34 & --33:37:28.0 & 17.82 & --1.41&  0.093  & 0.086 & 9   &    --0.358  & 0.084 & 17  & &       \\
  90018  & 00:59:05.64 & --33:34:42.0 & 18.91 & --2.59&  --0.097 & 0.087 & 8   &    --0.516  & 0.12  & 12  & &     \\
  90026  & 00:59:20.23 & --33:40:53.8 & 18.35 & --1.57&  0.039  & 0.068 & 11  &    --0.307  & 0.084 & 17  & &       \\
  90027  & 00:59:16.68 & --33:40:30.2 & 17.84 & --1.23&  0.302  & 0.087 & 9   &    --0.304  & 0.097 & 15  & &       \\
  90029  & 00:59:15.13 & --33:39:43.8 & 18.01 & --1.54&  0.18   & 0.08  & 10  &    --0.328  & 0.082 & 17  & &       \\
  90030  & 00:59:09.55 & --33:39:06.0 & 18.32 & --1.12&  0.266  & 0.113 & 7   &    --0.32   & 0.099 & 14  & &       \\
  90042  & 00:58:36.14 & --33:47:28.8 & 19.0  & --1.56&  --0.106 & 0.139 & 5   &    --0.324  & 0.114 & 13  & &     \\
  90045  & 00:58:33.33 & --33:47:47.8 & 18.83 & --1.46&  0.067  & 0.119 & 7   &    --0.295  & 0.114 & 13  & &       \\
  90050  & 00:57:58.22 & --33:43:51.5 & 18.66 & --2.52&  0.044  & 0.045 & 17  &    --0.445  & 0.043 & 32  & &       \\
  90057  & 00:58:02.10 & --33:39:29.5 & 19.37 & --2.03&  0.039  & 0.066 & 12  &    --0.461  & 0.062 & 22  & &       \\
  90069  & 01:01:39.57 & --33:42:15.9 & 18.12 & --1.72&  0.07   & 0.07  & 11  &    --0.349  & 0.066 & 22  & &       \\
  90070  & 01:01:41.03 & --33:38:44.1 & 18.88 & --1.62&  0.036  & 0.075 & 10  &    --0.333  & 0.077 & 19  & &       \\
  90074  & 01:01:45.83 & --33:37:10.7 & 19.24 & --2.0 &  0.074  & 0.074 & 11  &    --0.438  & 0.087 & 16  & &       \\
  90085  & 01:02:00.20 & --33:40:39.5 & 19.36 & --2.24&  0.682  & 0.139 & 7   &    --0.231  & 0.124 & 12  & &       \\
  90094  & 01:01:50.09 & --33:53:43.1 & 18.99 & --2.93&  --0.0   & 0.038 & 20  &    --0.444  & 0.046 & 30  & &     \\
  90096  & 01:01:36.27 & --33:47:01.1 & 18.24 & --2.51&  0.017  & 0.032 & 24  &    --0.421  & 0.028 & 49  & &       \\
  90104  & 01:01:19.26 & --33:45:41.7 & 17.51 & --1.75&  0.138  & 0.03  & 27  &    --0.343  & 0.032 & 45  & &       \\
  90109  & 01:01:18.42 & --33:42:47.0 & 19.33 & --2.11&  --0.039 & 0.034 & 23  &    --0.484  & 0.038 & 36  & &2    \\
  90112  & 01:01:08.55 & --33:45:35.0 & 17.15 & --1.56&  0.211  & 0.045 & 18  &    --0.318  & 0.03  & 47  & &       \\
  90118  & 01:00:49.37 & --33:42:00.6 & 17.05 & --1.92&  0.076  & 0.061 & 13  &    --0.382  & 0.073 & 19  & &       \\
  90121  & 01:01:12.54 & --33:41:03.9 & 17.93 & --1.7 &  0.019  & 0.033 & 24  &    --0.395  & 0.036 & 39  & &       \\
  90122  & 01:01:20.09 & --33:41:45.3 & 18.61 & --1.19&  0.131  & 0.083 & 13  &    --0.347  & 0.062 & 32  & &       \\
  90123  & 01:01:15.08 & --33:42:41.8 & 17.65 & --1.76&  0.041  & 0.044 & 25  &    --0.402  & 0.051 & 39  & &       \\
  90130  & 01:01:02.01 & --33:39:28.8 & 17.99 & --1.7 &  0.083  & 0.069 & 11  &    --0.342  & 0.087 & 16  & &       \\
  90131  & 01:00:54.17 & --33:40:14.6 & 17.13 & --1.67&  0.17   & 0.094 & 8   &    --0.314  & 0.105 & 14  & &       \\
  90325  & 00:59:08.56 & --33:36:43.9 & 18.7  & --1.53&  --0.045 & 0.066 & 11  &    --0.43   & 0.072 & 20  & &     \\
  90343  & 01:00:59.03 & --33:51:11.5 & 18.13 & --1.76&  0.092  & 0.07  & 11  &    --0.377  & 0.075 & 19  & &1       \\   
  90344  & 01:00:55.72 & --33:50:11.3 & 19.13 & --1.73&  0.146  & 0.082 & 10  &    --0.317  & 0.111 & 13  & &       \\   
  90354  & 01:01:11.51 & --33:51:16.6 & 19.61 & --1.69&  0.022  & 0.07  & 11  &    --0.317  & 0.083 & 17  & &       \\
  90355  & 01:01:05.39 & --33:49:23.9 & 19.9  & --1.71&  0.012  & 0.142 & 5   &    --0.345  & 0.114 & 13  & &       \\
  90361  & 01:01:09.30 & --33:52:34.6 & 18.49 & --1.55&  0.026  & 0.056 & 14  &    --0.364  & 0.089 & 16  & &       \\
  90365  & 01:01:45.00 & --33:51:39.0 & 19.73 & --2.0 &  0.021  & 0.07  & 11  &    --0.306  & 0.076 & 19  & &       \\
  90368  & 01:01:12.56 & --33:51:07.8 & 18.94 & --1.74&  0.087  & 0.09  & 9   &    --0.312  & 0.112 & 13  & &       \\
  90448  & 00:58:43.53 & --33:50:30.8 & 18.17 & --1.54&  0.161  & 0.061 & 13  &    --0.343  & 0.063 & 23  & &       \\
  90449  & 00:58:46.03 & --33:50:04.6 & 19.42 & --1.93&  0.0090 & 0.055 & 24  &    --0.4495 & 0.065 & 30  & &       \\
  90452  & 00:58:41.31 & --33:51:49.5 & 16.97 & --1.87&  0.103  & 0.057 & 14  &    --0.362  & 0.059 & 24  & &       \\
  90455  & 00:58:50.91 & --33:48:29.4 & 19.04 & --1.76&  0.0050 & 0.06  & 18  &    --0.325  & 0.056 & 35  & &       \\
  90457  & 00:58:42.60 & --33:48:33.4 & 16.93 & --1.92&  0.088  & 0.054 & 14  &    --0.351  & 0.065 & 22  & &       \\
  90458  & 00:59:20.65 & --33:48:56.6 & 17.05 & --1.73&  0.085  & 0.06  & 13  &    --0.328  & 0.042 & 34  & &       \\
  90549  & 01:00:26.29 & --33:44:45.7 & 18.0  & --1.62&  0.065  & 0.06  & 13  &    --0.362  & 0.058 & 25  & &       \\
  90552  & 01:00:50.87 & --33:45:05.2 & 17.16 & --1.49&  0.221  & 0.081 & 10  &    --0.292  & 0.101 & 14  & &       \\
  90553  & 01:00:42.50 & --33:44:23.5 & 17.11 & --1.35&  0.17   & 0.08  & 10  &    --0.312  & 0.095 & 15  & &       \\

\end{longtable}
\end{longtab}

\begin{longtab}
\setlength{\tabcolsep}{0.12cm}

\begin{longtable}{@{}lrrrrrrrrrrrrr}
\caption{Atmospheric parameters and C  and N  abundances for the sample stars.\label{tab_abundances}}\\

\hline \hline
ID            &  T    &  eT  &  $\log$(g)-l &  $\log$(g) &  $\log$(g)-u &  v$_{t}$  & ev$_{t}$ &  [O/Fe] &  [C/Fe]   &  [C/Fe]$_{\rm{cor}}$  & e[C/Fe]  &  [N/Fe] &  e[N/Fe] \\ 
               & (K)   &  (K) & (dex)          & (dex)         & (dex)          &   (km/s)   & (km/s)    &  (dex)    &  (dex)    &  (dex)                        & (dex)    & (dex)   & (dex)            \\ 
\hline
\endfirsthead
\caption{continued.}\\
\hline \hline
ID            &  T    &  eT  &  $\log$(g)-l &  $\log$(g) &  $\log$(g)-u &  v$_{t}$  & ev$_{t}$ &  [O/Fe] &  [C/Fe] &  [C/Fe]$_{\rm{cor}}$  &  e [C/Fe]  &  [N/Fe] &  e[N/Fe] \\ 
               & (K)   &  (K) & (dex)    & (dex)  & (dex)     &   (km/s)  & (km/s)    &  (dex)    &  (dex)    &  (dex) &  (dex)      & (dex)   & (dex)            \\ 
\hline
\endhead
\hline
\endfoot
  11       &   4589 & 31   &  1.0&        1.1   &   1.2   &     1.07  &  0.02 &    0.5   &  --0.5  &  0.2    &    0.16 & 0.59   & 0.31    \\ 
  12       &   4680 & 31   &  1.2&        1.4   &   1.4   &     1.09  &  0.01 &    0.48  &  --0.45 &  0.09   &    0.16 & 0.34   & 0.31     \\ 
  15       &   4589 & 31   &  1.0&        1.1   &   1.2   &     1.07  &  0.01 &    0.44  &  --0.56 &  0.14   &    0.15 & 0.14   & 0.29     \\ 
  17       &   4554 & 31   &  0.9&        1.0   &   1.0   &     1.05  &  0.01 &    0.5   &  --0.5  &  0.25   &    0.17 & 0.57   & 0.3      \\ 
  19       &   4512 & 31   &  0.8&        1.0   &   1.0   &     1.06  &  0.02 &    0.5   &  --0.49 &  0.25   &    0.16 & 0.46   & u.l.     \\ 
  20       &   4593 & 31   &  1.0&        1.1   &   1.1   &     1.06  &  0.01 &    0.5   &  --0.62 &  0.07   &    0.17 & 0.9    & 0.3      \\ 
  21       &   4635 & 31   &  1.1&        1.2   &   1.3   &     1.08  &  0.01 &    0.5   &  --0.49 &  0.19   &    0.16 & 0.06   & u.l.     \\ 
  22       &   4770 & 32   &  1.4&        1.5   &   1.5   &     1.1   &  0.01 &    0.5   &  --0.8  &  --0.35  &    0.17 & 0.3    & 0.28   \\ 
  25       &   4635 & 31   &  1.2&        1.3   &   1.4   &     1.09  &  0.01 &    0.4   &  --0.41 &  0.16   &    0.15 & 0.57   & 0.29     \\ 
  28       &   4409 & 31   &  0.7&        0.7   &   0.7   &     1.03  &  0.01 &    0.5   &  --0.67 &  0.1    &    0.17 & 0.42   & 0.34     \\ 
  30       &   4679 & 31   &  1.3&        1.3   &   1.5   &     1.09  &  0.01 &    0.36  &  --0.62 &  0.06   &    0.15 & 0.29   & 0.28     \\ 
  31       &   4740 & 32   &  1.3&        1.4   &   1.5   &     1.08  &  0.01 &    0.5   &  --0.45 &  0.12   &    0.17 & 0.2    & u.l.     \\ 
  32       &   4726 & 31   &  1.4&        1.5   &   1.6   &     1.1   &  0.01 &    0.4   &  --0.29 &  0.16   &    0.15 & 0.43   & 0.29     \\ 
  34       &   4647 & 31   &  1.1&        1.2   &   1.3   &     1.08  &  0.01 &    0.5   &  --0.88 &  --0.24  &    0.16 & 0.91   & 0.33   \\ 
  38       &   4648 & 31   &  1.1&        1.2   &   1.3   &     1.07  &  0.01 &    0.5   &  --0.7  &  --0.05  &    0.17 & 0.95   & 0.29   \\ 
  40       &   4878 & 32   &  1.6&        1.7   &   1.8   &     1.12  &  0.01 &    0.5   &  --0.75 &  --0.44  &    0.25 & 0.17   & 0.35   \\ 
  42       &   4673 & 31   &  1.2&        1.3   &   1.4   &     1.09  &  0.01 &    0.43  &  --0.82 &  --0.2   &    0.15 & 0.29   & 0.29   \\ 
  44       &   4631 & 31   &  1.2&        1.3   &   1.4   &     1.08  &  0.01 &    0.32  &  --0.55 &  0.0    &    0.14 & 0.32   & 0.27     \\ 
  46       &   4311 & 31   &  0.6&                0.8   &   0.8   &     1.1   &  0.02 &    0.31  &  --0.64 &  0.04   &    0.14 & 0.08   & 0.28        \\ 
  47       &   4673 & 66   &  1.2&        1.4   &   1.5   &     1.09  &  0.02 &    0.4   &  --0.67 &  --0.15  &    0.15 & 0.47   & 0.31   \\ 
  49       &   4681 & 32   &  1.3&        1.4   &   1.6   &     1.1   &  0.02 &    0.35  &  --0.58 &  --0.07  &    0.15 & 0.53   & 0.28   \\ 
  53       &   4621 & 31   &  1.1&        1.3   &   1.3   &     1.08  &  0.01 &    0.4   &  --0.9  &  --0.31  &    0.15 & 0.44   & 0.29   \\ 
  56       &   4727 & 32   &  1.3&        1.4   &   1.6   &     1.09  &  0.01 &    0.44  &  --0.25 &  0.26   &    0.16 & 0.75   & 0.3      \\ 
  57       &   4558 & 65   &  1.0&        1.2   &   1.3   &     1.08  &  0.03 &    0.42  &  --0.76 &  --0.12  &    0.15 & 0.55   & 0.29   \\ 
  60       &   4668 & 31   &  1.1&        1.2   &   1.3   &     1.07  &  0.01 &    0.5   &  --0.53 &  0.12   &    0.18 & 0.79   & 0.34     \\ 
  61       &   4481 & 31   &  0.8&        0.9   &   1.0   &     1.06  &  0.02 &    0.47  &  --0.35 &  0.35   &    0.15 & 0.72   & 0.28     \\ 
  107      &   4696 & 32   &  1.2&        1.4   &   1.4   &     1.09  &  0.01 &    0.5   &  --0.31 &  0.23   &    0.16 & 0.26   & 0.3      \\ 
  141      &   4638 & 31   &  1.2&        1.3   &   1.4   &     1.09  &  0.01 &    0.32  &  --0.3  &  0.23   &    0.14 & --0.09  & u.l.   \\ 
  155      &   4751 & 32   &  1.3&        1.4   &   1.5   &     1.08  &  0.01 &    0.5   &  --0.48 &  0.09   &    0.18 & 0.6    & 0.3      \\ 
  156      &   4956 & 33   &  1.8&        1.8   &   1.9   &     1.13  &  0.01 &    0.5   &  --0.17 &  0.06   &    0.18 & --0.09  & u.l.   \\ 
  158      &   4675 & 65   &  1.1&        1.2   &   1.4   &     1.07  &  0.02 &    0.5   &  --0.92 &  --0.21  &    0.21 & 0.25   & 0.4    \\ 
  182      &   4542 & 31   &  1.0&        1.1   &   1.3   &     1.08  &  0.02 &    0.36  &  --0.73 &  --0.07  &    0.14 & 0.25   & 0.29   \\ 
  183      &   4766 & 32   &  1.4&        1.6   &   1.6   &     1.11  &  0.01 &    0.39  &  --0.83 &  --0.42  &    0.15 & 0.1    & 0.3    \\ 
  197      &   4620 & 31   &  1.2&        1.4   &   1.4   &     1.1   &  0.01 &    0.27  &  --0.32 &  0.16   &    0.14 & 0.11   & 0.26     \\ 
  198      &   4384 & 31   &  0.7&        0.9   &   0.9   &     1.07  &  0.02 &    0.37  &  --0.53 &  0.18   &    0.15 & 0.0    & 0.29     \\ 
  206      &   4586 & 31   &  1.0&        1.1   &   1.2   &     1.07  &  0.02 &    0.5   &  --1.1  &  --0.49  &    0.16 & 0.94   & 0.29   \\ 
  208      &   4392 & 30   &  0.8&        0.9   &   1.0   &     1.09  &  0.02 &    0.28  &  --0.68 &  --0.01  &    0.14 & 0.19   & 0.26   \\ 
  210      &   4458 & 30   &  0.8&        1.0   &   1.0   &     1.07  &  0.01 &    0.38  &  --0.75 &  --0.06  &    0.15 & 0.48   & 0.27   \\ 
  220      &   4575 & 31   &  1.0&        1.2   &   1.2   &     1.08  &  0.01 &    0.43  &  --1.1  &  --0.54  &    0.15 & 1.06   & 0.28   \\ 
  10001    &   4588 & 31   &  1.0&        1.1   &   1.2   &     1.07  &  0.02 &    0.44  &  --0.45 &  0.22   &    0.15 & 0.16   & 0.29     \\ 
  10002    &   4406 & 31   &  0.8&        1.0   &   1.0   &     1.09  &  0.02 &    0.31  &  --0.66 &  0.01   &    0.14 & --0.28  & 0.27   \\ 
  10003    &   4842 & 32   &  1.5&        1.6   &   1.7   &     1.11  &  0.01 &    0.5   &  --1.15 &  --0.77  &    0.18 & 0.73   & 0.33   \\ 
  10004    &   4807 & 34   &  1.5&        1.7   &   1.8   &     1.11  &  0.01 &    0.41  &  --0.54 &  --0.21  &    0.15 & 0.77   & 0.3    \\ 
  20001    &   4601 & 31   &  1.1&        1.2   &   1.2   &     1.08  &  0.01 &    0.45  &  --0.38 &  0.25   &    0.15 & 0.23   & 0.27     \\ 
  20002    &   4631 & 66   &  1.1&        1.2   &   1.4   &     1.08  &  0.02 &    0.5   &  1.3   &  1.49   &    0.17 & 0.87   & 0.31      \\ 
  20003    &   4852 & 35   &  1.5&        1.6   &   1.7   &     1.11  &  0.01 &    0.5   &  0.12  &  0.49   &    0.18 & --0.22  & u.l.     \\ 
  20005    &   4676 & 31   &  1.2&        1.3   &   1.4   &     1.08  &  0.01 &    0.48  &  --1.03 &  --0.38  &    0.17 & --0.52  & u.l.  \\ 
  30001    &   4653 & 31   &  1.4&        1.5   &   1.6   &     1.11  &  0.01 &    0.19  &  --1.01 &  --0.55  &    0.13 & 0.3    & 0.26   \\ 
  30002    &   4865 & 32   &  1.7&        1.9   &   1.9   &     1.13  &  0.01 &    0.31  &  --0.09 &  0.04   &    0.15 & 0.42   & 0.33     \\ 
  50001    &   4607 & 31   &  1.2&        1.4   &   1.5   &     1.11  &  0.01 &    0.15  &  --0.29 &  0.13   &    0.13 & 0.4    & 0.35     \\ 
  50002    &   4499 & 31   &  0.8&        0.8   &   0.9   &     1.04  &  0.01 &    0.5   &  --1.05 &  0.43   &    0.19 & 0.78   & 0.35     \\ 
  50004    &   4704 & 65   &  1.2&        1.4   &   1.6   &     1.09  &  0.02 &    0.46  &  --0.39 &  0.14   &    0.17 & 0.47   & 0.32     \\ 
  90015    &   4292 & 30   &  0.8&        0.8   &   1.0   &     1.11  &  0.02 &    0.04  &  --0.9  &  --0.2   &    0.12 & 0.01   & 0.24   \\ 
  90018    &   4843 & 33   &  1.5&        1.6   &   1.7   &     1.1   &  0.01 &    0.5   &  --0.66 &  --0.3   &    0.24 & 0.46   & u.l.   \\ 
  90026    &   4507 & 31   &  1.1&        1.2   &   1.4   &     1.1   &  0.02 &    0.13  &  --0.69 &  --0.16  &    0.12 & 0.07   & 0.23   \\ 
  90027    &   4235 & 30   &  0.8&        0.9   &   1.0   &     1.14  &  0.02 &    -0.07 &  --0.84 &  --0.34  &    0.11 & --0.01  & 0.23  \\ 
  90029    &   4458 & 31   &  1.0&        1.1   &   1.2   &     1.1   &  0.02 &    0.11  &  --0.58 &  --0.03  &    0.12 & 0.18   & 0.23   \\ 
  90030    &   4351 & 31   &  1.0&        1.1   &   1.3   &     1.14  &  0.02 &    -0.13 &  --0.75 &  --0.31  &    0.24 & 0.03   & 0.35   \\ 
  90042    &   4640 & 32   &  1.4&        1.5   &   1.6   &     1.12  &  0.01 &    0.12  &  --0.41 &  0.01   &    0.13 & --0.05  & u.l.   \\ 
  90045    &   4543 & 31   &  1.2&        1.4   &   1.5   &     1.13  &  0.02 &    0.07  &  --0.46 &  --0.01  &    0.12 & --0.07  & 0.22  \\ 
  90050    &   4762 & 32   &  1.3&        1.4   &   1.5   &     1.09  &  0.01 &    0.5   &  --0.84 &  --0.28  &    0.18 & --0.09  & u.l.  \\ 
  90057    &   4980 & 37   &  1.9&        2.0   &   2.1   &     1.14  &  0.01 &    0.39  &  --0.63 &  --0.56  &    0.16 & 0.72   & 0.32   \\ 
  90069    &   4508 & 31   &  1.1&        1.1   &   1.3   &     1.09  &  0.01 &    0.21  &  --0.82 &  --0.13  &    0.13 & 0.17   & 0.25   \\ 
  90070    &   4609 & 31   &  1.3&        1.4   &   1.5   &     1.11  &  0.01 &    0.16  &  --0.76 &  --0.23  &    0.13 & 0.27   & 0.23   \\ 
  90074    &   4945 & 33   &  1.8&        2.0   &   2.1   &     1.14  &  0.01 &    0.37  &  --0.47 &  --0.4   &    0.16 & 1.0    & 0.31   \\ 
  90085    &   5073 & 35   &  2.1&        2.2   &   2.3   &     1.16  &  0.01 &    0.5   &  1.35  &  1.38   &    0.15 & 1.59   & 0.25      \\ 
  90094    &   4815 & 34   &  1.5&        1.5   &   1.6   &     1.1   &  0.01 &    0.5   &  --0.07 &  0.36   &    0.2  & 1.01   & 0.35     \\ 
  90096    &   4530 & 31   &  0.8&        0.9   &   1.0   &     1.04  &  0.01 &    0.5   &  --0.71 &  0.06   &    0.18 & 0.56   & 0.35     \\ 
  90104    &   4263 & 30   &  0.6&        0.6   &   0.8   &     1.08  &  0.02 &    0.23  &  --0.9  &  0.17   &    0.13 & 0.03   & 0.28     \\ 
  90109    &   4939 & 35   &  1.8&        1.9   &   2.1   &     1.13  &  0.01 &    0.44  &  --0.87 &  --0.74  &    0.16 & 0.3    & u.l.   \\ 
  90112    &   4022 & 30   &  0.5&        0.6   &   0.6   &     1.17  &  0.02 &    0.12  &  --1.02 &  --0.26  &    0.16 & --0.35  & u.l.  \\ 
  90118    &   4068 & 30   &  0.5&        0.5   &   0.6   &     1.14  &  0.02 &    0.33  &  --0.68 &  --0.01  &    0.15 & 0.45   & 0.27   \\ 
  90121    &   4446 & 31   &  0.9&        1.0   &   1.1   &     1.08  &  0.02 &    0.2   &  --1.23 &  --0.54  &    0.13 & 0.28   & 0.25   \\ 
  90122    &   4462 & 31   &  1.2&        1.3   &   1.4   &     1.13  &  0.01 &    -0.09 &  --1.18 &  --0.6   &    0.11 & --0.22  & 0.23  \\ 
  90123    &   4353 & 31   &  0.8&        0.8   &   0.9   &     1.08  &  0.02 &    0.24  &  --1.14 &  --0.4   &    0.13 & 0.11   & 0.26   \\ 
  90130    &   4448 & 31   &  0.9&        1.0   &   1.1   &     1.08  &  0.02 &    0.2   &  --0.83 &  --0.21  &    0.13 & 0.04   & 0.25   \\ 
  90131    &   4093 & 30   &  0.6&        0.6   &   0.6   &     1.13  &  0.02 &    0.19  &  --1.01 &  --0.28  &    0.13 & --0.23  & 0.3   \\ 
  90325    &   4826 & 31   &  1.7&        1.8   &   2.0   &     1.13  &  0.01 &    0.11  &  --1.0  &  --0.7   &    0.13 & 0.12   & 0.28   \\ 
  90343    &   4490 & 31   &  1.0&        1.1   &   1.2   &     1.08  &  0.02 &    0.24  &  --0.97 &  --0.27  &    0.13 & 0.17   & 0.25   \\ 
  90344    &   4622 & 65   &  1.3&        1.3   &   1.5   &     1.09  &  0.02 &    0.22  &  --0.28 &  0.21   &    0.16 & 0.1    & 0.3      \\ 
  90354    &   4709 & 65   &  1.4&        1.6   &   1.7   &     1.12  &  0.02 &    0.2   &  --0.55 &  --0.2   &    0.16 & --0.26  & 0.3   \\ 
  90355    &   4861 & 37   &  1.7&        1.8   &   2.0   &     1.13  &  0.01 &    0.21  &  --0.46 &  --0.22  &    0.14 & 0.01   & u.l.   \\ 
  90361    &   4632 & 31   &  1.3&        1.4   &   1.5   &     1.11  &  0.01 &    0.12  &  --1.17 &  --0.61  &    0.13 & 0.03   & 0.23   \\ 
  90365    &   4900 & 36   &  1.7&        1.9   &   2.0   &     1.13  &  0.01 &    0.37  &  --0.1  &  0.01   &    0.15 & --0.31  & 0.3    \\ 
  90368    &   4621 & 31   &  1.3&        1.3   &   1.5   &     1.09  &  0.02 &    0.23  &  --0.42 &  0.1    &    0.13 & 0.13   & 0.23     \\ 
  90448    &   4437 & 31   &  1.0&        1.1   &   1.3   &     1.1   &  0.02 &    0.11  &  --0.76 &  --0.21  &    0.12 & --0.09  & 0.23  \\ 
  90449    &   4929 & 34   &  1.8&        1.9   &   2.0   &     1.14  &  0.01 &    0.33  &  --0.76 &  --0.61  &    0.15 & 0.22   & 0.31   \\ 
  90452    &   4179 & 30   &  0.5&        0.6   &   0.6   &     1.1   &  0.02 &    0.3   &  --0.87 &  --0.12  &    0.14 & --0.13  & 0.3   \\ 
  90455    &   4683 & 32   &  1.4&        1.4   &   1.6   &     1.1   &  0.01 &    0.24  &  --0.6  &  --0.11  &    0.14 & 0.08   & 0.25   \\ 
  90457    &   4139 & 30   &  0.5&        0.5   &   0.6   &     1.1   &  0.01 &    0.33  &  --0.87 &  --0.13  &    0.14 & 0.45   & 0.32   \\ 
  90458    &   4089 & 30   &  0.5&        0.5   &   0.5   &     1.13  &  0.02 &    0.22  &  --1.06 &  --0.33  &    0.14 & 0.15   & 0.31   \\ 
  90549    &   4435 & 31   &  1.0&        1.0   &   1.1   &     1.1   &  0.01 &    0.16  &  --0.94 &  --0.23  &    0.13 & 0.15   & 0.24   \\ 
  90552    &   3906 & 30   &  0.4&        0.4   &   0.5   &     1.2   &  0.02 &    0.08  &  --0.85 &  --0.28  &    0.13 & 0.07   & 0.3    \\ 
  90553    &   4032 & 30   &  0.4&        0.4   &   0.5   &     1.14  &  0.02 &    0.0   &  --1.13 &  0.0    &    0.12 & 0.05   & 0.27     \\ 

\end{longtable}
\end{longtab}

\subsection{Stellar atmosphere parameters}
For the measurement of elemental abundances, we need to know the star's effective temperature ( \teff), 
gravity ($\log$(g)), microturbolent velocity (v$_{\rm mic}$), and iron abundance [Fe/H]. In addition, the determination of the 
C and N abundance is dependent upon the assumed [O/Fe] abundance ratio.

In the following, we assume as [Fe/H] values and their measurement errors 
those derived in previous spectroscopic studies of the 
Sculptor dSph using the semi-empirical calibrations between the equivalent widths of the near-infrared Ca~II triplet
lines and [Fe/H] tested over the 
range --4 $<$ [Fe/H] $<$ --0.5 dex  by \citet{starkenburg10}. 

The photometric \teff~was derived from the \citet{ramirez05} 
color-metallicity-\teff~relations for giant stars, 
adopting the polynomial fit corrections appropriate to the metallicity of each given star. For most stars, we determine 
\teff~as a weighted average of the temperature derived from the \vi, \vj, \vh, \vk~relations, \teffc. 
The photometry in $V$ and $I$ band 
comes from deep, wide-area CTIO/MOSAIC~II data by \citet{deboer11}, while the infrared $JHKs$
photometry is from VISTA/VIKING survey data. Of the 94 targets, eight 
were not present in the \citet{deboer11} data since they 
probably fell onto bad pixels or were too close to saturated stars; for these, 
we used previous ESO/WFI $V$ \& $I$ photometric data \citep{tolstoy04}. The photometry was de-reddened using the 
\citet{schlegel98} maps.

For the \teff~errors in each of the individual color-metallicity-\teff~relations, $\sigma_{\rm{T}\rm{eff,col}}$,  
we sum in quadrature three terms: the first term is derived from the standard error propagation 
considering the measurement uncertainties in [Fe/H], and in the $V$ and $I$ photometry; the second term is the scatter 
found by \citet{ramirez05} around the T$_{VI}$-[Fe/H] color relations; and the third term is the scatter 
between direct temperature (i.e. stellar angular diameter and bolometric flux measurements) 
and those derived by the infrared flux method (50K; see \citealp{ramirez05b}). 
The final error we quote for \teff~is the error in the weighted average, i.e. 
$\sigma_{\rm{Teff}} = (\sum{1/\sigma_{\rm{Teff,col}}^2})^{-1/2}$. 
For three stars, we could only use the \vi~color, 
as these stars were not present in the available infrared photometric data set. 
The $\sigma_{\rm{Teff}} $ for these stars amounts to $\simeq$60 K, as compared to $\simeq$30 K for the rest of the sample.

\begin{figure}
\begin{center}
\includegraphics[width=6cm]{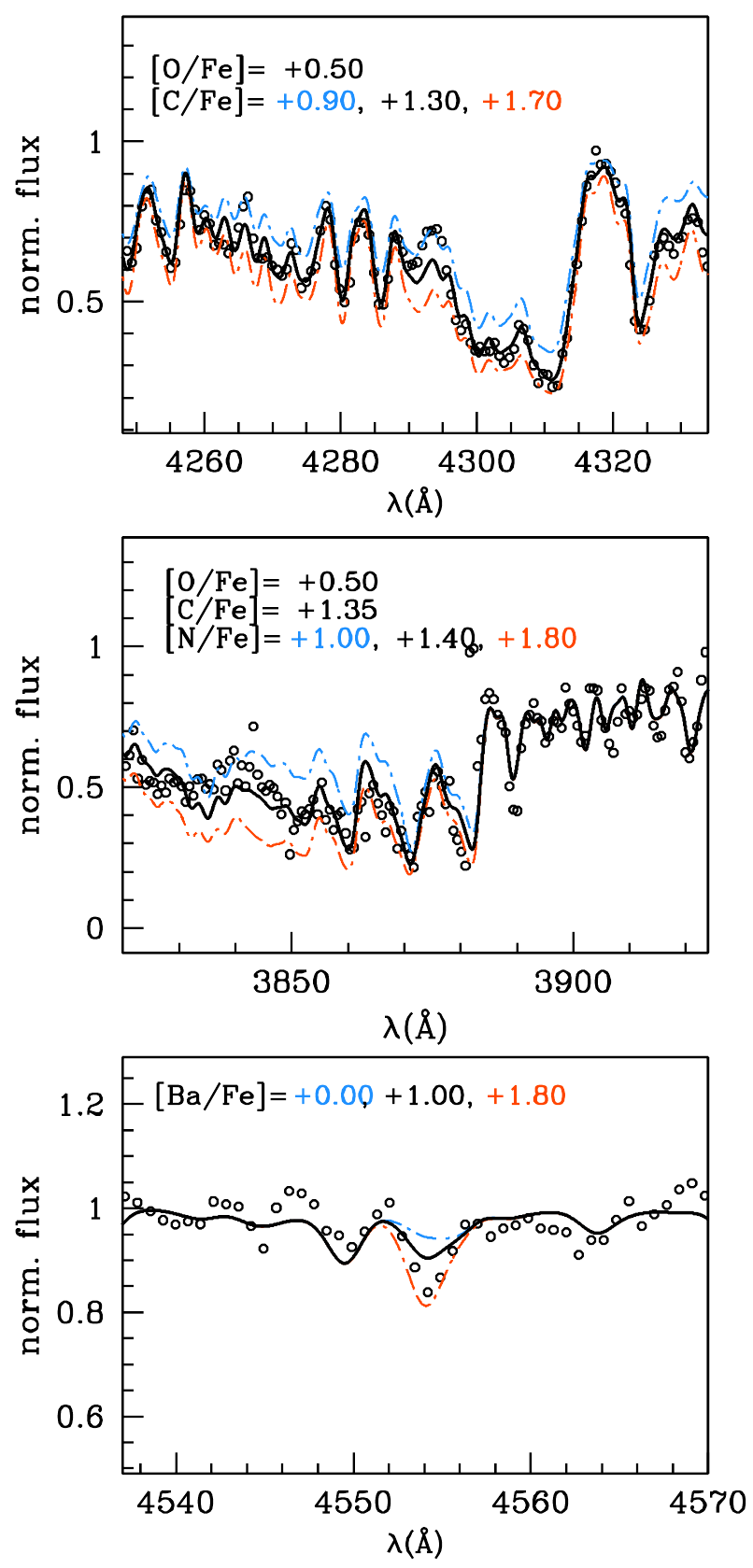}
\caption{Observed (black dots) and synthetic (blue, black, and red lines)  spectra around the CH (upper),
CN (middle), and Ba (bottom)  features  band for the star 90085 ($V$=19.36, SNR=7 and 12 at 3880 and 4300\AA\,, respectively) 
to illustrate the quality of the VIMOS spectra.}
\label{fig:SPECTRA}
\end{center}
\end{figure}

The $\log$(g) was determined by finding the point along the RGB locus of a set of 
Dartmouth isochrones\footnote{\url{http://stellar.dartmouth.edu/models/index.html}} with the 
closest \teff, [Fe/H], [$\alpha$/Fe] and age to the values corresponding to a given Sculptor star. We use 
Dartmouth isochrones of --2.4 $\le$ [Fe/H] $\le$ --0.6 dex and $-0.2 \le$ [$\alpha$/Fe] $\le +0.6$ dex, both spaced of 0.2 dex. 
For our sample of Sculptor stars, we assume the trend in [$\alpha$/Fe] as a function of [Fe/H] observed in 
a high-resolution FLAMES/GIRAFFE pointing (\citealp{tolstoy09}, Hill et al. in prep.) located in the 
Sculptor central regions, corresponding to [$\alpha$/Fe]= 0.33 dex at [Fe/H]$< -1.87$ and 
[$\alpha$/Fe] = --0.5~$\times$~[Fe/H] -- 0.605 dex at larger [Fe/H], and we adopt an error in [$\alpha$/Fe] of $\pm$0.1dex. 
The underlying assumption is that  [$\alpha$/Fe] as a function of [Fe/H] does not vary across the 
galaxy. This kind of assumption appears to hold for 
the dSphs for which this has been verified so far, i.e., Fornax  \citep{letarte10,hendricks14,lemasle14}. 
All the Sculptor stars 
have [$\alpha$/Fe] within the isochrone range, while only two are more metal-poor than [Fe/H]=--2.4 dex outside of their 
1$\sigma$~error-bars. \citet{deboer12} assigned  likely ages to the Sculptor RGB stars with spectroscopic observations, on the basis of the 
star formation history they derived from deep photometry. This yields ages older than 8 Gyr for the majority of 
Sculptor stars. We then assigned to the stars in our sample either an age of 8 Gyr or an age of 12 Gyr, according to 
the closest value from \citet{deboer12} . 
The errors in $\log$(g) were 
determined by repeating the fit 100 times, where in each of these 100 Montecarlo realisations the [Fe/H], \teff~and 
[$\alpha$/Fe] values were 
randomly extracted from a Gaussian distribution centred on observed values 
and with standard deviation equal to the measurement error. 

For the microturbolent velocity, we used the formula obtained by the GES survey (\citealp{gilmore12}, Bergemann et al., in preparation),
which depends on \teff, $\log$(g), and metallicity. Here again the errors on v$_{\rm mic}$ were obtained by standard error propagation. 

 Table \ref{tab_abundances} reports the \teff, $\log$(g) and v$_{\rm mic}$ values and their 
 uncertainties; these values are used to choose the atmospheric model for the spectral synthesis of each star.

\begin{figure}
\begin{center}
\includegraphics[width=6.0cm]{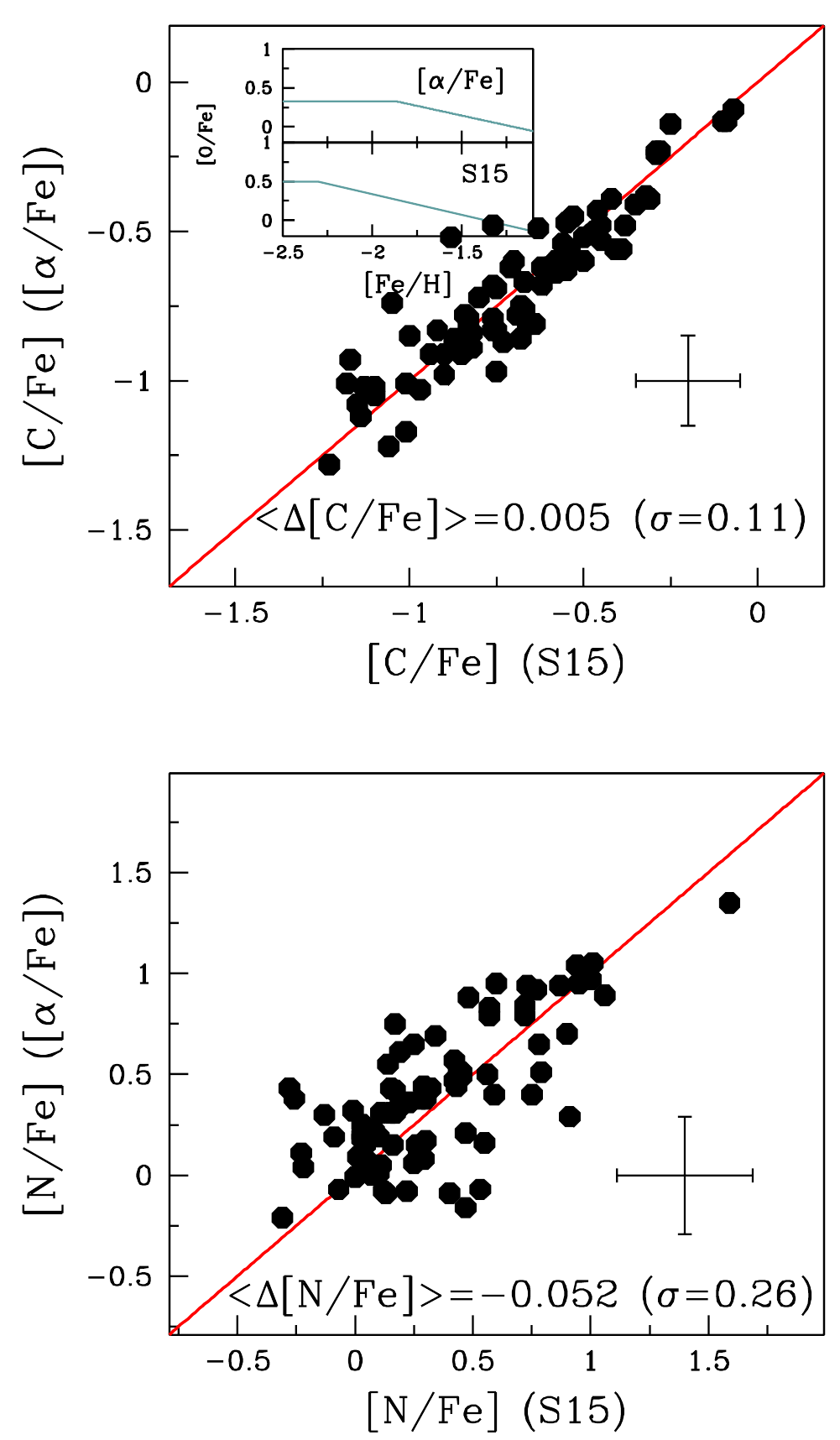}
\caption{[C/Fe] and [N/Fe] abundances computed by setting 
  [O/Fe] = [$\alpha$/Fe] (denoted with the [$\alpha$/Fe] label; see inset)
  are compared to [C/Fe] abundances obtained by assuming the oxygen trend reported in \citet{skuladottir15}, denoted with the S15 label.
  Each panel lists the mean difference with its $\sigma$ in the measured abundances.
  The red line represents the 1:1 relation.}
\label{fig:OSSIGENO}
\end{center}
\end{figure}

\begin{figure*}
\begin{center}
\includegraphics[width=17.5cm]{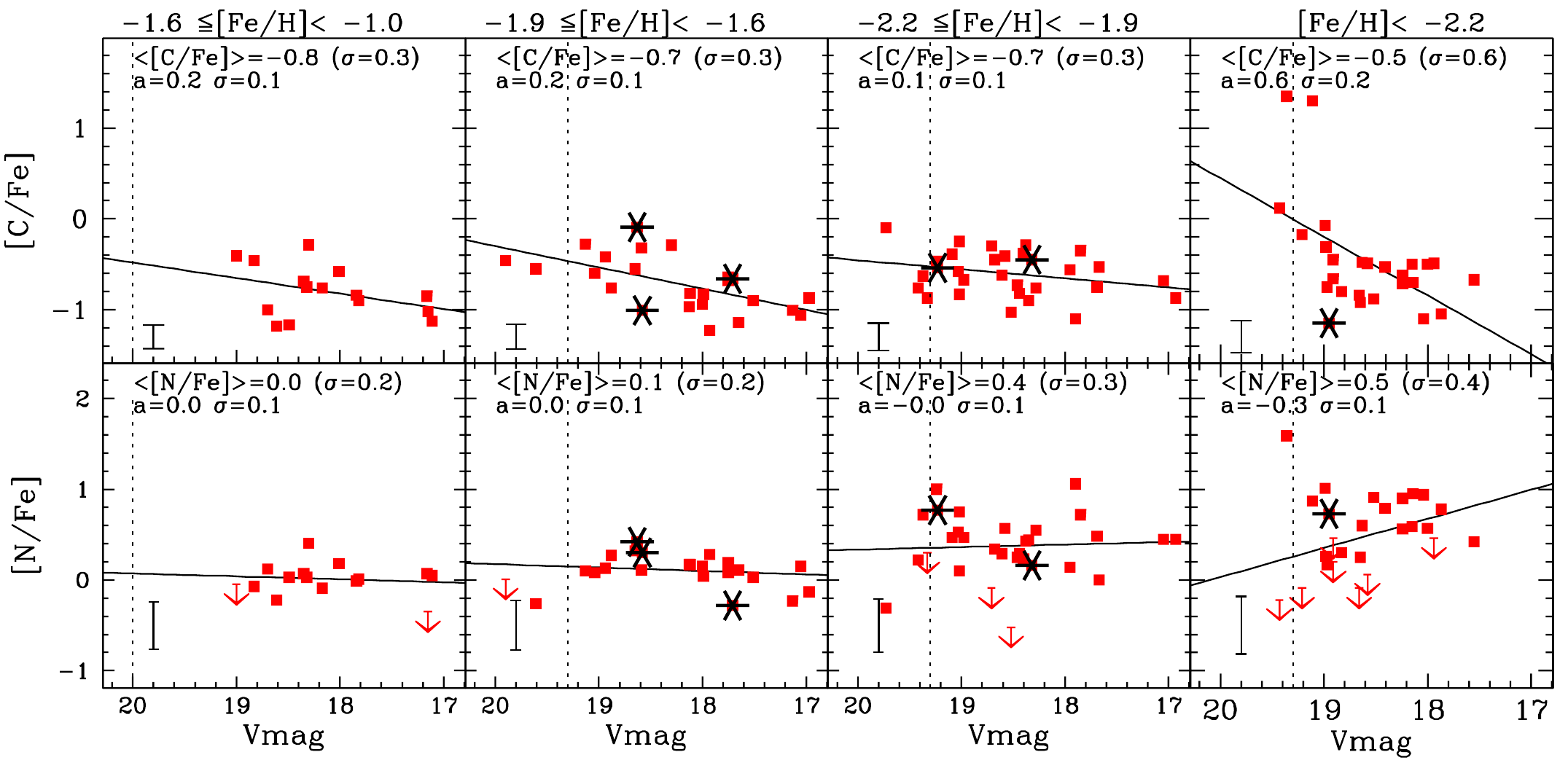}
\caption{Run of the [C/Fe] and [N/Fe] abundance ratio versus extinction-corrected 
visual magnitude in four bins of metallicity for Sculptor stars, with the 
black asterisks denoting objects belonging to the kinematic substructure. 
The vertical line indicates the location of the RGB bump (where deep mixing onsets) for the metal-poor 
and metal-rich component of Sculptor (see \citealp{majewski99}).}
\label{fig:ABUNDANCES}
\end{center}
\end{figure*}
\subsection{Spectral synthesis}
We compared the observed spectra with template synthetic spectra to derive quantitative estimates of C and N 
abundances. The atomic and molecular line lists were taken from the latest Kurucz compilation \citep{castelli04} and 
downloaded from F. Castelli's website\footnote{\url{http://wwwuser.oat.ts.astro.it/castelli/linelists.html}} 
both for atomic and molecular transitions. Model atmospheres were calculated with the ATLAS9 code 
starting from the grid of models available in F. Castelli's website \citep{castelli03}, 
using the appropriate  values of [Fe/H],~\teff, $\log$(g), and v$_{\rm mic}$ as derived in the previous section.
The ATLAS9 models employed were computed with the new set of opacity distribution functions
\citep{castelli03} and excluded approximate overshooting in 
calculating the convective flux. 

C and N abundances were estimated via spectral synthesis of the CH band at
$\simeq$4300~\AA~and the CN bands at $\simeq$3883~\AA, respectively.
We computed model spectra  by means of the SYNTHE code developed by Kurucz.
We derived abundances  through a $\chi^{2}$minimization between the observed spectrum and a grid of synthetic spectra calculated at different abundances. The synthetic spectra have been convolved to the instrumental resolution, then rebinned at the same pixel step as the observed spectra. The upper and middle panels of Fig.~\ref{fig:SPECTRA} illustrate the fit of synthetic spectra to the observed spectra in CH and CN spectral regions. Abundances for C and N were determined together in an iterative way because for the temperature of our stars, carbon and nitrogen form molecules and, as a consequence, their abundances are related to each other. Reference solar abundances are from \citet{asplund09}.

To estimate the sensitivity of the derived A(C) and A(N) abundances\footnote{A(C) and A(N) are the carbon and 
nitrogen abundances in the usual logarithmic scale, i.e. A(X)=$\log$(N$_{\rm{X}}$/N$_{\rm{H}}$)+12.}
 to the adopted atmospheric parameters, 
we repeated our abundance analysis by changing only one parameter at each iteration (using the errors 
given in Table~\ref{tab_abundances}) for several stars that are representative of the temperature and gravity range explored.
Typically, for the temperature we find $\delta$A(C) / $\delta$ \teff $\simeq$ 0.05 dex  
and $\delta$A(N) / $\delta$ \teff $\simeq$ 0.06 dex. We measure 
$\delta$A(C) / $\delta$ $\log$(g) $\simeq$ 0.03 dex  
and $\delta$A(N) / $\delta$ $\log$(g) $\simeq$ 0.06 dex for the surface gravity, and $\delta$A(C) / $\delta$ v$_{\rm mic}$ $\simeq$ 0.02 dex  
and $\delta$A(N) / $\delta$ v$_{\rm mic}$ $\simeq$ 0.03 dex for the microturbulent velocity.

The derived abundances are also dependent on the assumed oxygen abundance. 
Throughout the analysis, we assume the [O/Fe] versus [Fe/H] trend adopted by \citet{skuladottir15} based on the trend observed for a subset of the stars in the 
high-resolution FLAMES/GIRAFFE pointing by \citet{tolstoy09} and Hill et al. in prep. 
We  also explored a slightly different trend, where [O/Fe] =  [$\alpha$/Fe]; as shown in Fig.~\ref{fig:OSSIGENO} 
this would have had little impact on the final 
carbon and nitrogen determination, causing a star-to-star random change of about 0.005 dex. 
As a conservative quantification of the sensitivity of the C abundance on the adopted O abundance, 
we varied the oxygen abundances by as much as 0.4 dex 
and repeated the spectrum synthesis to determine the exact dependence for the same three representative stars 
(4100 K $\leq$ \teff~$\leq$ 4900 K). 
We find that changes in the oxygen abundance of 0.4 dex 
cause a variation in the derived [C/Fe]  of $\delta$A(C) / [O/Fe] $\simeq$ 0.08 dex.
This error was then included in the final error computation on [C/Fe] and [N/Fe]. 

In addition to the stellar parameters and oxygen abundance errors, a measurement uncertainty exists in 
 the determination of the individual abundances. This intrinsic error 
 was estimated by means of Monte Carlo simulations and combined with the general errors above.
Briefly, we repeated the fitting procedure using a sample of 1000 synthetic spectra, 
where Poissonian noise has been injected (after the re-mapping of the synthetic spectra 
at the same pixel step of the VIMOS spectra)  to reproduce the noise conditions observed around the analysed bands. 
These uncertainties are of the order of $\simeq$0.09 and $\simeq$ 0.23 dex for  C and N, respectively.
All these individual errors were added in quadrature and gave a final error 
$\Delta$[C/Fe] = 0.14 dex, $\Delta$[N/Fe] = 0.30 dex.

All the above individual error sources and abundances are listed in Table~\ref{tab_abundances}. 
We present the abundances derived as described above and the relative uncertainties in Table~\ref{tab_abundances}. 

\citet{skuladottir15} estimate [C/Fe] upper limits for 85 Sculptor stars,
using VLT/FLAMES intermediate resolution data from the CN molecular line at  9100-9250~\AA.
There are nine stars in our program for which  \citet{skuladottir15} report abundances.
The latter derive C estimates and upper limits on the C abundance from the CN band, assuming 
 [C/N]=--1.2 (i.e. the average value for mixed stars in the range --3 $\leq$[Fe/H] $\leq$ --2 dex; \citealp{spite05}) and [N/Fe]=0.0 dex;
 e.g., [C/Fe]$_{ext}$ and [C/Fe]$_{lim}$ of their Table~6, respectively.
We rederived carbon abundances from VIMOS spectra in the CN in the 3880\AA~ region, under 
the same assumptions as \citet{skuladottir15}, to compare the measured abundances. 
The comparison is satisfying, with the data scattering about the line corresponding to equality. 
In most cases, our C abundances are very close to those by \citet{skuladottir15}, 
mostly within  0.1~dex, with a spread on the whole sample of 0.4~dex. 
In the case of star 210,  carbon abundances have also been measured from medium-resolution spectra by \citet{kirby15}. We
 measure for this star [C/Fe]=--0.75 $\pm$ 0.15 dex, which is somewhat higher than the carbon abundances reported by 
 \citet{kirby15} and \citet{skuladottir15}, [C/Fe]=--0.91 and --1.01, respectively, but still within
 1-1.5$\sigma$\footnote{To compare \citet{kirby15} and \citet{skuladottir15} abundances to our data, we shift their abundances to our adopted solar scale \citep{asplund09}.}.

\section{Results}\label{RESULTS}

\subsection{C and N abundances and mixing along the RGB}\label{MIXING}
The gathered data set makes it possible 
for the first time to  analyse the C and N abundances trends simultaneously in this galaxy over a 
wide range in [Fe/H], and explore the effects of internal mixing along the RGB for stars 
above the RGB bump.  

In Fig.~\ref{fig:ABUNDANCES} we plot the [C/Fe] and [N/Fe] abundance ratios for all the observed stars as a function of their $V$ magnitude.
We distinguish between four subsamples according to the star's
metallicity (see Sect.~\ref{IND}).  
Each panel reports the average [C/Fe] and [N/Fe] abundances (together with their dispersions), the slope, and the error of the linear fit.
Also indicated are the different locations of the RGB bump for the metal-poor and metal-rich populations of Sculptor \citep{majewski99}. 

Figure~\ref{fig:ABUNDANCES} shows the decline of [C/Fe] with advancing evolution along the RGB \citep[e.g.,][]{shetrone13,kirby15}. 
The luminosity-dependent trend detected in Fig.~\ref{fig:ABUNDANCES} is indicative 
of some deep mixing, which circulates material from the base of the convective envelope down into the CN(O)-burning region as soon as the star is brighter than the RGB bump \citep[e.g.,][]{charbonnel98,spite05}.
In this sense, the Sculptor stars appear to be behaving in a manner analogous to giants in metal-poor GCs \citep[e.g.,][]{martell08, shetrone10,kirby15}, as well as halo field giants of comparable metallicity  \citep{gratton00}. 
The carbon depletion appears to be more severe in the bin with the lowest metallicity, i.e. [Fe/H] < --2.2 dex. 
This is in agreement with \citet{martell08}, who find 
that the carbon depletion as a function of the time that the star has spent in the deep-mixing phase 
is approximately twice as large at $[{\rm Fe/H}]=-2.0$ than at --1.0 dex.

In dredged-up material processed by the CNO cycle, N is produced at the expense of C. Thus the atmosphere of the star should become N-rich and C-poor relative to its initial composition. On the contrary,  the decline in carbon is not accompanied by an increase in [N/Fe] abundance with luminosity (lower panels of Fig.~\ref{fig:ABUNDANCES}). The notable exception to this general trend concerns the stars with [Fe/H] < --2.2 dex, among which there is a trend of increasing  [N/Fe] with evolving state along the RGB. 
\begin{figure}
\begin{center}
\includegraphics[width=\columnwidth]{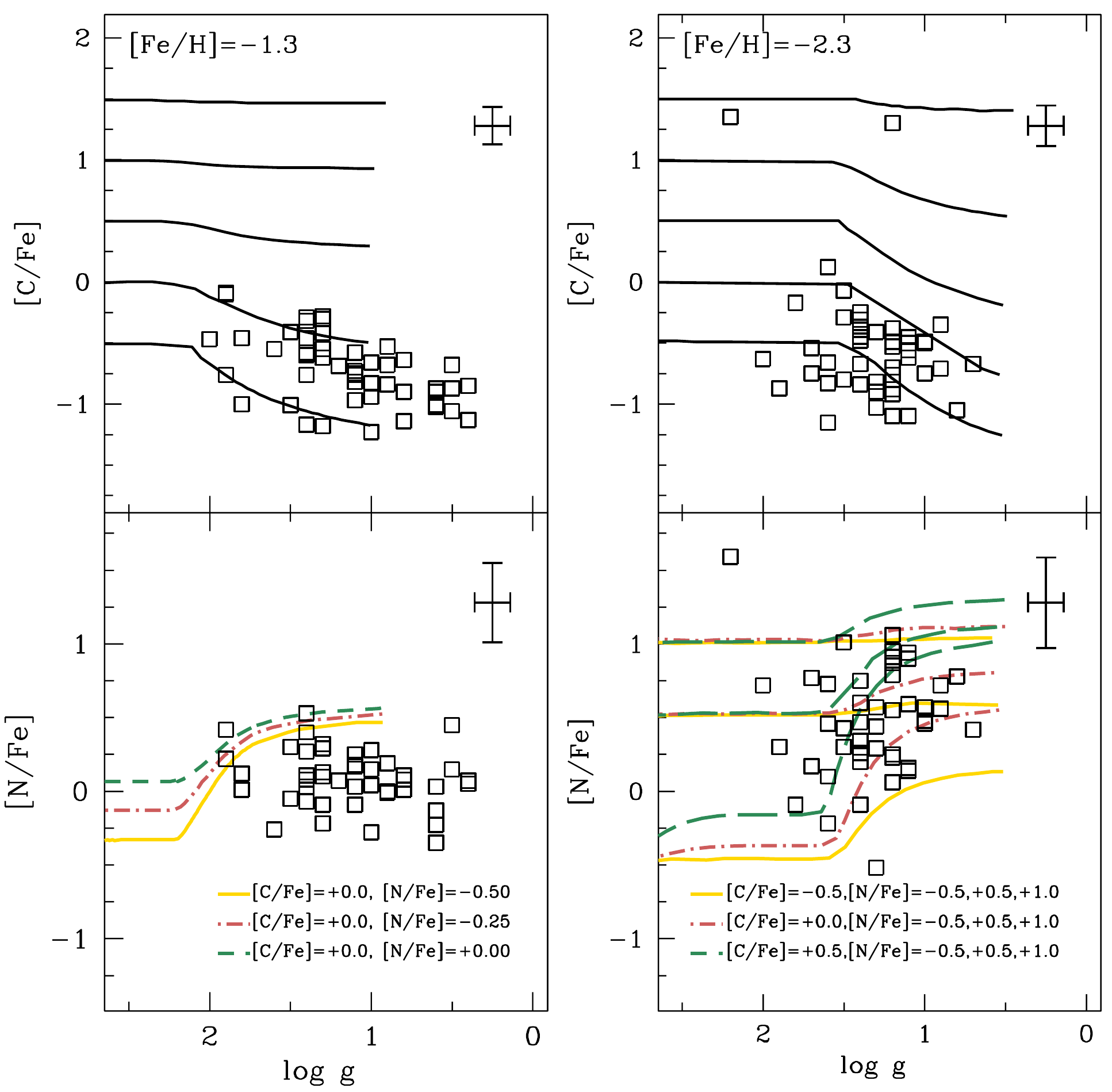}
\caption{[C/Fe] (top) and [N/Fe] (bottom) abundance ratios for stars
with metallicity [Fe/H] > --2.0 dex  (left-hand panel) and [Fe/H] $\leq$ --2.0 (right-hand panel) are compared with the theoretical models by \citet{placco14} at metallicity [Fe/H] =--1.3 and [Fe/H] =--2.3 dex.
The mean error on the abundances and the photometric gravity is shown in the top right corner of each panel.
Black solid lines (top panel only) are the theoretical models for [C/Fe] =1.5, 1.0, 0.5, 0.0, -0.5 dex  assuming [N/Fe]=+0.0.
Colored lines (bottom panels only) are the theoretical models for [N/Fe]. The color coding refers to the initial composition of the models; see the keys on the bottom of each plot.}
\label{fig:PLACCO}
\end{center}
\end{figure}

\begin{figure}
\begin{center}
\includegraphics[width=0.8\columnwidth]{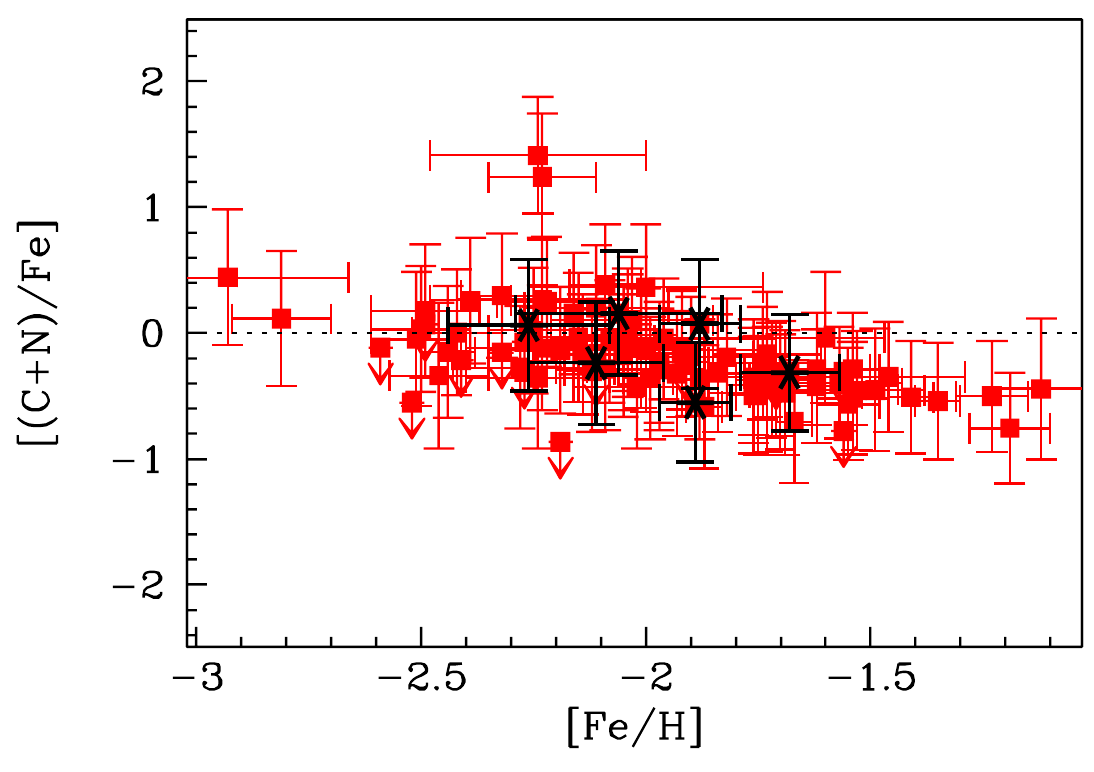}
\caption{[(C+N)/Fe] vs. [Fe/H] for the spectroscopic sample; symbols are as in Fig.~\ref{fig:CMD}.
There is a decline of  [(C+N)/Fe] with metallicity, an effect previously reported by \citet{spite06}.}
\label{fig:CNSUM}
\end{center}
\end{figure}

To further investigate the behaviour of C and N with RGB evolution, 
we compare the observed  [C/Fe] and [N/Fe] abundance ratios with the theoretical predictions for evolutionary 
mixing by \citet{placco14} in Fig.~\ref{fig:PLACCO}.
The top panels of Fig.~\ref{fig:PLACCO} show the measured carbon abundances 
for stars with metallicity [Fe/H] >--2.0 and $\leq$ --2.0 dex (left- and right-hand panel, respectively), along with the 
\citet{placco14} predictions for metallicity [Fe/H]=--1.3 and --2.3 dex and [C/Fe]= --0.5,+0.0,+0.5,+1.0,+1.5 dex, assuming [N/Fe]=+0.0 dex\footnote{The Placco et al. models of different [N/Fe] predict a very similar behaviour for [C/Fe] as a function of $\log$(g), except for very large [N/Fe]$=$+2, values that are
absent in our data.}.  From this plot, it is apparent that the behaviour of the [C/Fe] ratio with the evolutionary stage is perfectly reproduced by the theoretical predictions of 
\citet{placco14} for initial carbon abundance  --0.5 $\leq$ [C/Fe] $\leq$ 0.0 dex.

The bottom panels of Fig.~\ref{fig:PLACCO} show the same comparison, but for [N/Fe] abundances. 
The left-hand panel shows models for a metallicity of [Fe/H]=--1.3 and [C/Fe]= 0.0 dex 
and three different N abundances. Although the bulk of stars in this wide metallicity bin is significantly 
metal-poorer than [Fe/H]=--1.3 dex and is at lower [C/Fe] abundances (<[Fe/H]> $\simeq$--1.7 and <[C/Fe]> $\simeq$--0.7 dex, respectively)\footnote{Qualitatively, we expect that ,
for metal poor stars, the onset of mixing would be shifted to lower surface gravities and that less initial C leads to 
smaller N variations on the upper RGB (see for example Fig.~1 in \citealp{placco14}).}, 
theoretical models do not predict a large [N/Fe] variation in the surface gravity range explored by VIMOS data.
Moreover, the increase of [N/Fe] with luminosity, if any, would have been hidden by the large error on [N/Fe] abundances.
This quantitatively explains why we do not observe in Fig.~\ref{fig:ABUNDANCES} a trend of [N/Fe] abundances with luminosity
for stars more metal-rich than [Fe/H] $\simeq$ --2.0 dex.
In the bottom, right-hand panel, \citet{placco14} models for  [Fe/H]=--2.3 dex, [C/Fe]=--0.5,+0.0,+0.5 dex, and initial N abundance [N/Fe]=--0.5,+0.5,+1.0 dex are compared with [N/Fe] for stars with [Fe/H]$\leq$ --2.0 dex.  From this figure, one can see that the measured N abundances for stars with [Fe/H] $\leq$ --2.0 dex is well reproduced by theoretical models with [C/Fe]=-0.5, +0.0, +0.5 dex and metallicity [Fe/H]=--2.3 dex.
At lower metallicities, \citet{placco14} models foresee a larger N-enhancement with advancing 
evolution. This is consistent with the observed trend of [N/Fe] vs. Vmag observed in Fig.~\ref{fig:ABUNDANCES} for the most metal-poor stars
in our sample.

Finally, \citet{spite06} find that the [(C+N)/Fe] ratio is slightly larger for metal-poor stars than for metal-rich stars (see their Fig.~10).
This effect is not expected under the assumption that any excess of N results only from the transformation of C into N, maybe 
indicating some metallicity dependence in the efficiency of mixing. Figure~\ref{fig:CNSUM} shows that a trend of [(C+N)/Fe] with metallicity is also emerging from our data. However, in stars with this luminosity, carbon is highly depleted and the total number of C+N atoms is close to the number of N atoms. Therefore, the observed trend of Fig.~\ref{fig:CNSUM} could also be explained by systematic errors in the N abundance determinations \citep[NLTE, 3D-effects; e.g.,][]{spite05,spite06}.

\subsection{Evolution of C and N abundances}\label{EVOLUTION}

The average values of [C/Fe] in the four metallicity ranges of Fig.~\ref{fig:ABUNDANCES} 
suggest that the stars with higher metallicity started
 with lower [C/Fe] ratios than stars with [Fe/H] $\leq$--2.20 dex \citep[see also][]{kirby15}. 
We recover the initial carbon abundance ([C/Fe]$_{\rm{cor}}$) for each star in our sample by applying the corrections 
computed by \citet{placco14}, on the basis of models of mixing on the upper RGB that  depend on surface gravity, metallicity, and uncorrected carbon abundances. The intent is to explore to what extent the observed trend of [C/Fe] with metallicity is due to star-to-star intrinsic
variation in carbon abundances rather than to 
stellar evolution. 
\begin{figure}
\begin{center}
\includegraphics[width=0.9\columnwidth]{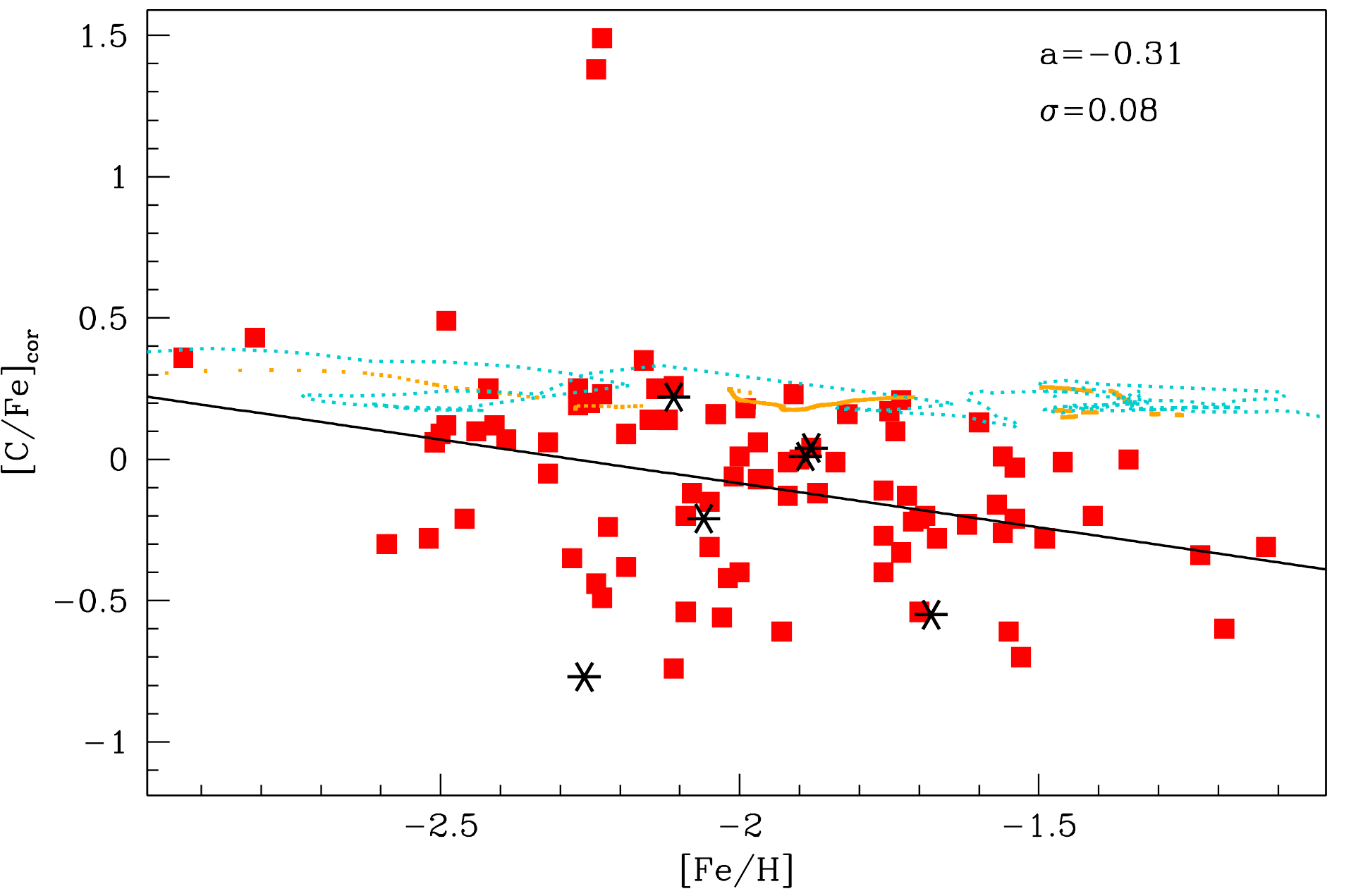}
\caption{[C/Fe] abundances corrected for luminosity according to \citet{placco14} as a function of stellar metallicity.
  The black solid line is the weighted linear fit to the data.
  The slope ($a$) of the fit and the rms ($\sigma$) are also indicated.
  The dotted lines are the model presented by \citet{romano13} (see text).}
\label{fig:CCORRETTO}
\end{center}
\end{figure}

\begin{figure}
\begin{center}
\includegraphics[width=0.9\columnwidth]{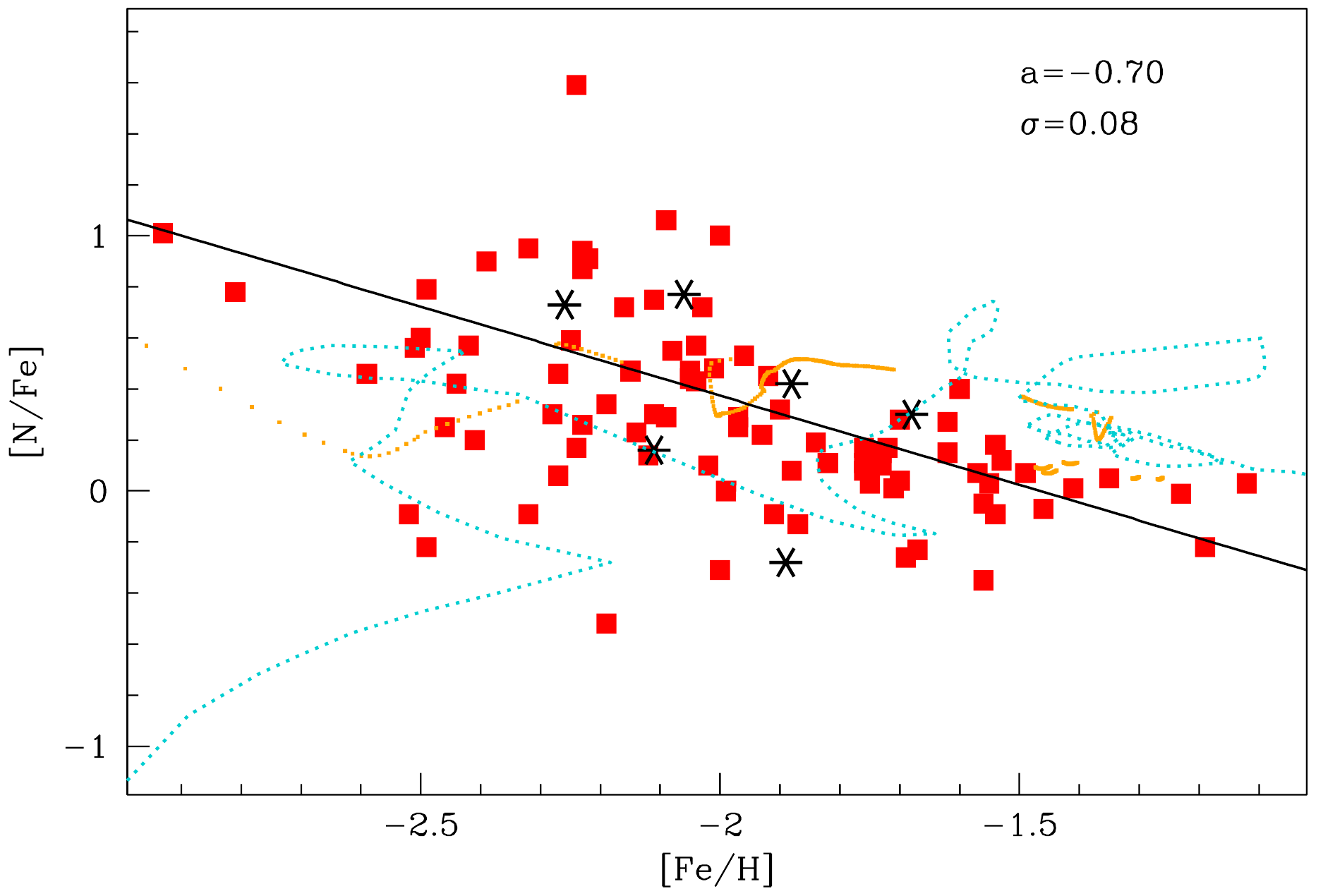}
\caption{The same as Fig.~\ref{fig:CCORRETTO} but for [N/Fe] abundances.}
\label{fig:NDONATELLA}
\end{center}
\end{figure}

Figure~\ref{fig:CCORRETTO} confirms that the decline of [C/Fe] with increasing [Fe/H] in Sculptor cannot be entirely ascribed to stellar 
evolution. 
The observed decline in [C/Fe]$_{\rm{cor}}$  with metallicity can be interpreted by the dependence of the stellar yields on metallicity 
together with the growing contribution of SN~Ia  to the interstellar medium enrichment \citep{kirby15}.
The orange and cyan lines in Fig.~\ref{fig:CCORRETTO}  show the runs of [C/Fe] with metallicity predicted by models Scl2 and Scl4 of \citet{romano13}, respectively. These authors compute the detailed chemical evolution of Sculptor-like galaxies selected from the mock catalogue of Milky Way satellites by \citet{starkenburg2013}, by taking into account the IMF-averaged \citep{salpeter55}
 yields from different nucleosynthetic sources (AGB stars, SNeII and SNeIa), as well as the detailed stellar lifetimes (i.e. the instantaneous recycling approximation is relaxed). More specifically, the adopted stellar yields are from \citet{vandenhoek97} for low- and intermediate-mass stars, \citet{woosley95} for massive stars, and \citet{iwamoto99} for SNeIa. The star formation in the models is not continuous, but proceeds in bursts. When star formation is not active, we do not plot the corresponding values of the abundance ratios in the interstellar medium, since no stars form with that chemical composition; this is why the model predictions are indicated as broken lines.  Figure~\ref{fig:CCORRETTO}
shows that  the upper envelope of the data (when excluding the two C-rich stars)
 is reasonably well fitted by the models. It is worth emphasising that the goodness of the fit obviously depends on the goodness of the adopted yields. Chemical evolution models using the very same set of yields tend to slightly overestimate the [C/Fe] ratios of solar neighbourhood stars in the same metallicity range \citep{romano10}. Thus, the result for Sculptor is not totally unexpected. It is also worth stressing that the Sculptor models discussed here already nicely reproduce other observed features of this galaxy, such as the [Mg/Fe]--[Fe/H] relation and its stellar metallicity distribution (see \citealp{romano13})

The [C/Fe]$_{\rm{cor}}$ abundance ratios of  Fig.~\ref{fig:CCORRETTO} 
show a dispersion beyond the measurement uncertainty. When we fit a linear relation between
[C/Fe]$_{\rm{cor}}$ and [Fe/H], calculate the residuals about the fit, and divide 
them by the measurement uncertainties, we find $\Delta$=[C/Fe]$_{\rm{cor}}$/[C/Fe]$_{\rm{cor}}$(fit)/$\delta$[C/Fe]= 2.1
(for comparison, \citealp{kirby15} find 1.5). However,
 if the observed scatter were due to measurement uncertainties, the value of $\Delta$ would be 1. 
Stars in dSph galaxies originate from gas polluted by high-energy supernova ejecta and material from AGB stars of different masses and metallicities, and AGB stars can still pollute the interstellar medium while the galaxy is forming stars. Consequently, the observed residual scatter in carbon abundances can be due to enrichment by various nucleosynthetic sources over time of an interstellar medium, which is not well mixed \citep{kirby15}. As pointed out by \citet{kirby15}, there are several  other possible sources for
this residual scatter, including inhomogeneous mixing, underestimation of the error on the C abundances,
and small errors on the \citet{placco14} corrections.

The lower panel of Fig.~\ref{fig:ABUNDANCES} seems to show that the mean value of
[N/Fe] decreases with increasing metallicity, perhaps
reflecting a decrease in N production relative to Fe. 
In principle, the observed [N/Fe] decline can be attributed to the low S/N ratio of our spectra.
However, a plot of [N/Fe] abundances vs. SNR (not shown) shows that the 
upper limits on [N/Fe] are distributed somewhat evenly in the [N/Fe]-SNR plane, indicating the presence of a real trend. 
The [N/Fe] values show large scatter and our sample is composed of stars whose surface abundances have been altered by internal mixing. Furthermore, 
unlike for C-, we cannot access corrections to recover the initial [N/Fe], prior to the modifications induced by the mixing.  
The abundance trends are therefore difficult to verify and interpret.
However, from the comparison of the observed [N/Fe] ratios with the predictions from models Scl2 and Scl4 by \citet{romano13} for Sculptor
presented in Fig.~\ref{fig:NDONATELLA}, a general agreement can be found. The model forming stars less efficiently \citep[model Scl2; see][their Fig.~2 and Table~1]{romano13} does predict a complex behaviour: the [N/Fe] ratio goes up and down, and [Fe/H] back and forth, as a consequence of the various episodes of gas accretion onto or loss of gas from the galaxy.
The adopted nucleosynthetic yields do not include any primary N production from massive stars. Thus, according to the models, most of the nitrogen seen in Sculptor would be secondary in nature, coming mostly from intermediate-mass stars that experience hot bottom burning.

\begin{figure}
\begin{center}
\includegraphics[width=\columnwidth]{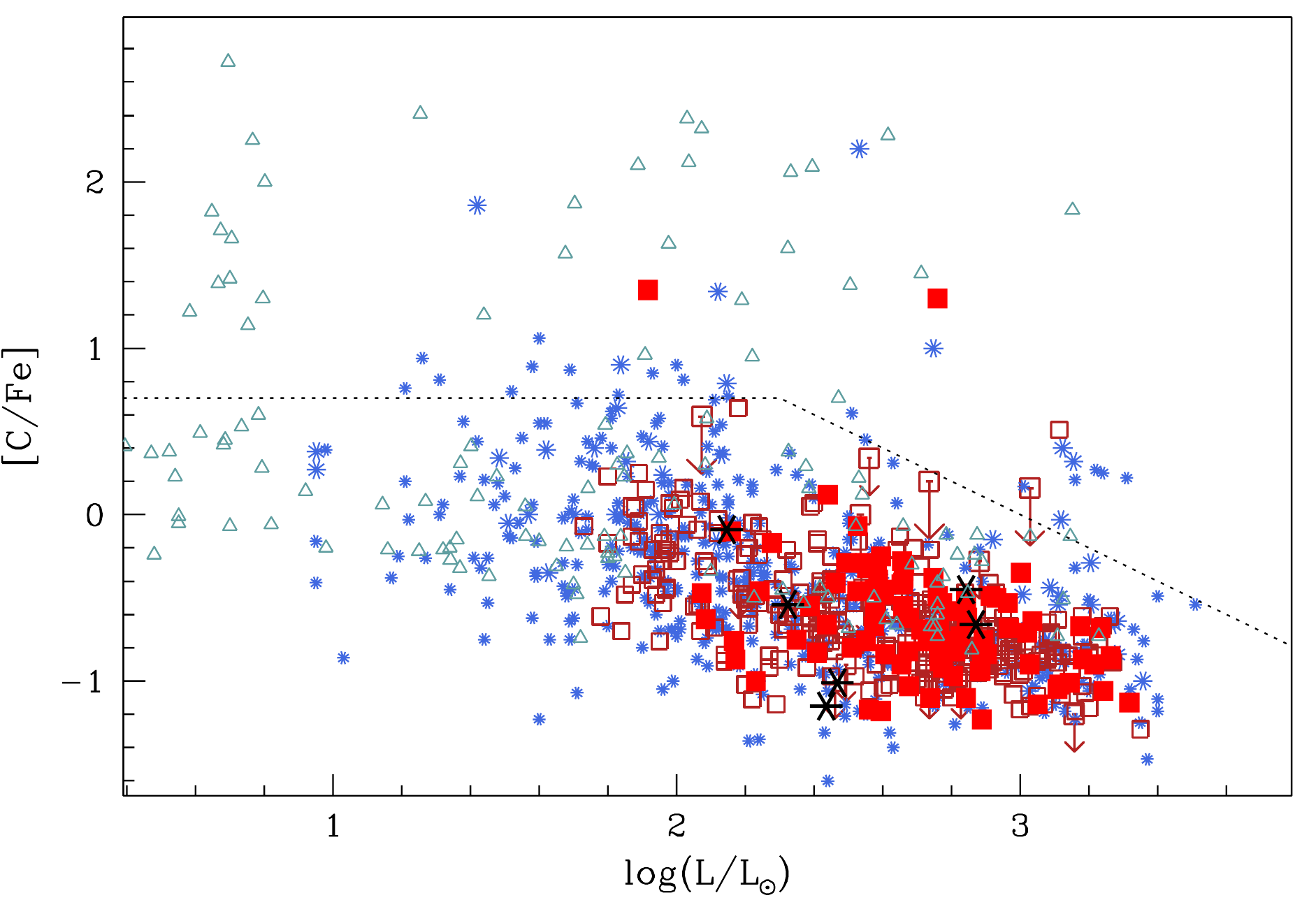}
\caption{[C/Fe] as a function of log(L/L$_{\sun}$) for Galactic dwarf satellite and halo red giants.
The Sculptor stars we analysed  are represented by red squares and black asterisks. Burgundy squares denote 
the [C/Fe] abundances of Sculptor stars from the literature \citep{tafelmeyer10,frebel10, starkenburg13,skuladottir15,kirby15,simon15}. Small light blue asterisks are [C/Fe] measurements for dSph stars from literature data  \citep{fulbright04,cohenhuang09, tafelmeyer10,cohenhuang10,norris10,lai11,honda11,venn12}. 
Empty green triangles represent the halo stars of \citet{yong13} with [Fe/H] $\geq$ --3.0 dex.
The dotted line is the \citet{aoki07} dividing line for carbon enhancement, which takes into account
the depletion of carbon with evolution along the RGB.}
\label{fig:CALIT}
\end{center}
\end{figure}

\subsection{CEMP stars and the connection with the Galactic halo}\label{CEMP}
In the MW halo a significant fraction of metal-poor stars, i.e., [Fe/H] $\leq$ --2 dex, 
is enriched in carbon ([C/Fe] $\geq$ 0.7 dex)\footnote{Throughout this paper, we adopt the \citet{aoki07} criterion, to define {carbon-enhanced} objects.}.
The fraction of CEMP stars also increases with decreasing metallicity \citep{beers05},
suggesting that large amounts of carbon were synthesised in the early Universe when the oldest and most metal-poor stars formed.
Traditionally, CEMP stars are  divided into two broad categories, according to their chemical composition:
carbon enriched stars that display an excess of heavy elements formed in slow ($s$) or rapid ($r$)
neutron capture processes (CEMP-$s$, CEMP-$r$, and CEMP-$r/s$)\footnote{Barium and europium are often taken as representatives 
of the $s$- and $r$-processes. In the solar system, the main component of the $s$-process contributes $\approx$82\% of the Ba, whereas $\approx$18\% is due to $r$-process \citep{arlandini99}. On the contrary, the $r$-process accounts for the production of 
$\approx$94\% Eu.  While low mass (M $\leq$ 4 M\sun) AGB stars are the stellar sources for the
$s$-process  \citep{gallino98,arlandini99,travaglio99}, the $r$-process requires a high-energy, 
neutron-rich environment, i.e., supernovae (e.g., \citealp{travaglio04}) and/or neutron 
star mergers \citep{tsujimoto14}.}; and CEMP-no, stars that display no 
such overabundance of neutron-capture elements.
The observed chemical pattern of CEMP-$s$  and CEMP-$r/s$  stars is thought to be the result of mass 
transfer in a binary system, coming from an evolved star in the AGB evolutionary stage
\citep[e.g.,][]{lucatello05,starkenburg2014}.
In contrast, both CEMP-no  and CEMP-$r$ more likely reflect the chemical pattern of the
interstellar medium out of which they formed
\citep[e.g.,][]{norris97,barklem05,aoki07}. The significantly distinct binary properties of CEMP-no and CEMP-s
stars strongly supports the hypothesis of a different physical origin of their carbon enhancements \citep{starkenburg2014}.

 
\begin{table}
\caption{CEMP star frequencies.}
\label{tab:FREQ}
\centering
\begin{tabular}{c l c c }
\hline\hline
Fe/H] $\leq$    &  \multicolumn{3}{l}{\% CEMP stars} \\
                          &   Scl dSph        &    Dwarfs  &    MW Halo \\
\hline
--2.0           &         2 (3) &     8  &  43 \\
--2.5           &        0 (4) &   22  &  44 \\
--3.0           &        0 (8) &   40  &  50 \\
--3.5           &        0 (17) &   50  &  62 \\
\hline  
\end{tabular}
\tablefoot{Fraction of CEMP stars (percent; according the \citealp{aoki07} criterion) per metallicity bins for Scl dSph; dwarf galaxies in the Local Group; and the Galactic halo. The same literature data are  plotted in Fig.~\ref{fig:CALIT}. The values in parenthesis refer to the fraction of CEMP stars in Scl when taking the [C/Fe] upper limit from \citet{frebel10} into account for star S102549, with [Fe/H]=--3.8 dex and [C/Fe]= +0.16 dex. }
\end{table}


Carbon measurements have been reported for a significant number of stars in Sculptor, and the available measurements, irrespective of the CEMP star classification, are shown in Fig.~\ref{fig:CALIT} together with [C/Fe] determinations from literature for MW dwarf satellites and halo red giants as a function of log(L/L$_{\odot}$). Our [C/Fe] determinations nicely fall in the region occupied by the other large samples available in literature \citep{skuladottir15,kirby15}.
The great majority of our sample is found to have [C/Fe] < 0 dex, analogous to that observed by \citet{kirby15}
for evolved stars in GCs (see for example their Fig.~6). Out of our 94 stars with [C/Fe] measurements, we find two stars 
(2002 and 90085) that are CEMP, i.e., the two CN-strong stars of Sect.~\ref{IND}, with around $\simeq$1.0 dex higher carbon abundance than other stars of similar luminosity or metallicity ([C/Fe] = 1.3 and 1.35 dex; [C/Fe]$_{\rm{cor}}$=1.49 and 1.38 dex, respectively). These are also among the most metal-poor stars in the VIMOS sample, with [Fe/H] $\simeq$ --2.2 dex.
Figure~\ref{fig:CALIT} also illustrates that so far, in Sculptor, only one other star has been shown to be a CEMP star\footnote{If we exclude the upper limit [C/Fe] measurement by  \citet{frebel10} just above the \citet{aoki07} criterion.}. This is the CEMP-no star ET0097 ([Ba/Fe] < 0 dex), with metallicity [Fe/H] = --2.03 $\pm$ 0.10, which is enhanced in carbon and nitrogen by [C/Fe]= 0.51 $\pm$ 0.10 and  [N/Fe] = 1.18 $\pm$ 0.20, respectively \citep{skuladottir15}. 

To investigate the nature of our two CEMP stars, we estimate their barium abundances, 
as the strong Ba line at 4554~\AA~is an 
accessible indicator of the $s$-processes. Although the low resolution combined with the relatively low SNR of our spectra
prevent us from measuring [Ba/Fe] abundances,  both spectra are compatible with having [Ba/Fe] > +1 dex (see Fig.~\ref{fig:SPECTRA}). This finding indicates that both stars might be part of a binary system with a companion star that went through the AGB phase causing pollution with material rich in carbon and $s$-process elements. 
Then, even though to-date [C/Fe] measurements exist for 369 stars, there still remains only one CEMP-no star in Sculptor. 
 
If the stellar halo of our Galaxy was 
formed long ago out of the shredded stellar component of satellite galaxies resembling the progenitors
of MW dSphs, then we would expect 
the oldest stars in both stellar systems to share the 
same abundance distributions.  
Currently, there are 118 stars in Sculptor, at metallicity [Fe/H] $\leq$ --2 dex,  with measurements, or upper limits, of [C/Fe] (\citealp{frebel10,tafelmeyer10,starkenburg13,skuladottir15,kirby15,simon15}; and this work). In the Galactic halo,
the proportion of CEMP stars of all classes is $\approx$40\% (\citealp{yong13}; see Table~\ref{tab:FREQ}).
If this fraction was the same in Sculptor, 
the expected number of CEMP stars with [C/Fe] $\geq$ 0.7 dex would be 47 $\pm$ 5.
However, only three CEMP stars have been found, with only the S15 being a CEMP-no star. 

As pointed out by \citet{skuladottir15}, the comparison with the MW halo becomes much more favourable
when focusing on the metallicity range -2.5 $\lesssim$ [Fe/H] $\lesssim$ -2, where the percentage of CEMP-no halo stars
decreases to 5$^{+3}_{-2}$. With our sample we add new measurements for other 41 stars in this [Fe/H] bin;
including the samples from S15 and K15, this adds up to 90
stars with measured [C/Fe], of which 3 CEMP stars including only 1 CEMP-no. Therefore in Sculptor, the
fraction of CEMP (CEMP-no) stars is 3.4\%. 

Very recently, \citet{salvadori15} used a statistical, data-calibrated cosmological model to investigate the apparent 
discrepancy between the observed frequency of CEMP stars in the Galactic halo and Sculptor.
They find that the fraction of CEMP stars as a function of [Fe/H] strongly depends
on the luminosity of the dwarf galaxy, but with a fraction approaching unity at [Fe/H]$\lesssim -4$ for all
objects. For a Sculptor-like galaxy though, whose metallicity distribution function is
dominated by stars with [Fe/H]$\simeq$-2, the joint probability of {\it observing} a star with a given
[Fe/H] that is also carbon enhanced is maximum in the range -2.5 $\lesssim$ [Fe/H] $\lesssim$ -2, being P= 0.02.
According to this probability, in a sample of 88 stars we should have observed 1.8$\pm$1.3 CEMP stars, which is still
in agreement with our findings. 

 \subsection{A disrupted globular cluster?}\label{DIS_GC}
 In this section we focus on the chemodynamical substructure found by \citet{battaglia07} in the Sculptor dSph and
 use our VLT/VIMOS data to investigate the possibility this may be the remnant of a disrupted globular cluster via
 searching for the distinctive C-N pattern of second-generation globular cluster stars. 

Anticorrelations between [C/Fe] and [N/Fe] are typically detected in GC
stars, even in unevolved stars, supposedly not affected by deep mixing \citep[e.g.,][]{cohen05}.
Such anticorrelations can also be
 detected at low spectral resolution using the
 CH and CN molecular bands as proxies for the C- and N- abundance, respectively, where stars
 can be divided in CN-strong/weak, CH-strong/weak groups or show a continuous spread of values 
 \citep[e.g.,][]{pancino10}.
 Second-generation GC stars are typically identified with the CN-strong/CH-weak groups. 
 As discussed in the previous sections, abundances of C and N in luminous giants are
significantly modified by mixing. Furthermore, in our case, we are examining a galactic environment,
where the stars have an intrinsic spread in 
metallicity and age. Since the effects of mixing vary also with metallicity, 
we would be able to conclude that stars with high N- and low C- belonging to the Sculptor substructure are second population GC stars if (at least) 
a subsample of stars in the substructure will show the strong CN and enhanced N abundances 
{relative to the rest of the data set over a similar range of metallicity, color, and magnitude}. 
This line of investigation was successfully adopted by \citet{martell10} and \citet{martell11}
on SEGUE spectra of MW halo stars, finding that 
3\% of the observed field MW halo stars show the chemical characteristics of GC stars. 

From Figs.~\ref{fig:INDEX} and~\ref{fig:ABUNDANCES}, we observe that none of the stars in the kinematic substructure displays 
either strong S($\lambda$3839) band strength or large [N/Fe] abundances with respect to stars with similar metallicity 
and magnitude. The CN band strength is very sensitive to the overall heavy element abundance in the atmosphere,
decreasing in abundance roughly proportionally to  the decrease in metallicity squared. 
Thus, in the metal-poor regime, possible metallicity effects may be masking 
the C and N signatures readily found in metal-rich GC stars (e.g., \citealp{briley93}, \citealp{smolinski11}).
However,  C and N inhomogeneities have been detected in 
GCs with metallicity comparable to the kinematic substructure in Scl dSph, both in stars fainter and
brighter than the RGB bump. In M~15 ([Fe/H]=--2.37 dex), stars with luminosities above the RGB bump exhibit in general weak S($\lambda$3883) CN bands \citep{trefzger83}, with a handful of stars having enhanced CN absorption \citep{langer92,lee00}. \citet{cohen05} find anticorrelated ranges of C and N abundances in unvolved RGB stars in the same cluster, with large star-to-star differences in [N/Fe] abundance ratio up to $\simeq$2.5 dex. \citet{norris85} and \citet{smolinski11} detect a CN bimodality among stars in M~92 ([Fe/H]=--2.31 dex), although the separation of the populations is minimal, as observed in other metal-poor clusters ( i.e., NGC 5053; see \citealp{smolinski11}). \citet{shetrone10} present CN band measurements for 67 stars (both below and up the RGB bump) in NGC 5466 ([Fe/H]=--1.98 dex). Their data suggest the presence of a CN bimodality, although the mean separation in CN band strength between the two groups is not large relative to the scatter within each group (see also \citealp{martell08b} in the case of M~55, with [Fe/H]=--1.94 dex). 

\begin{figure}
\begin{center}
\includegraphics[width=\columnwidth]{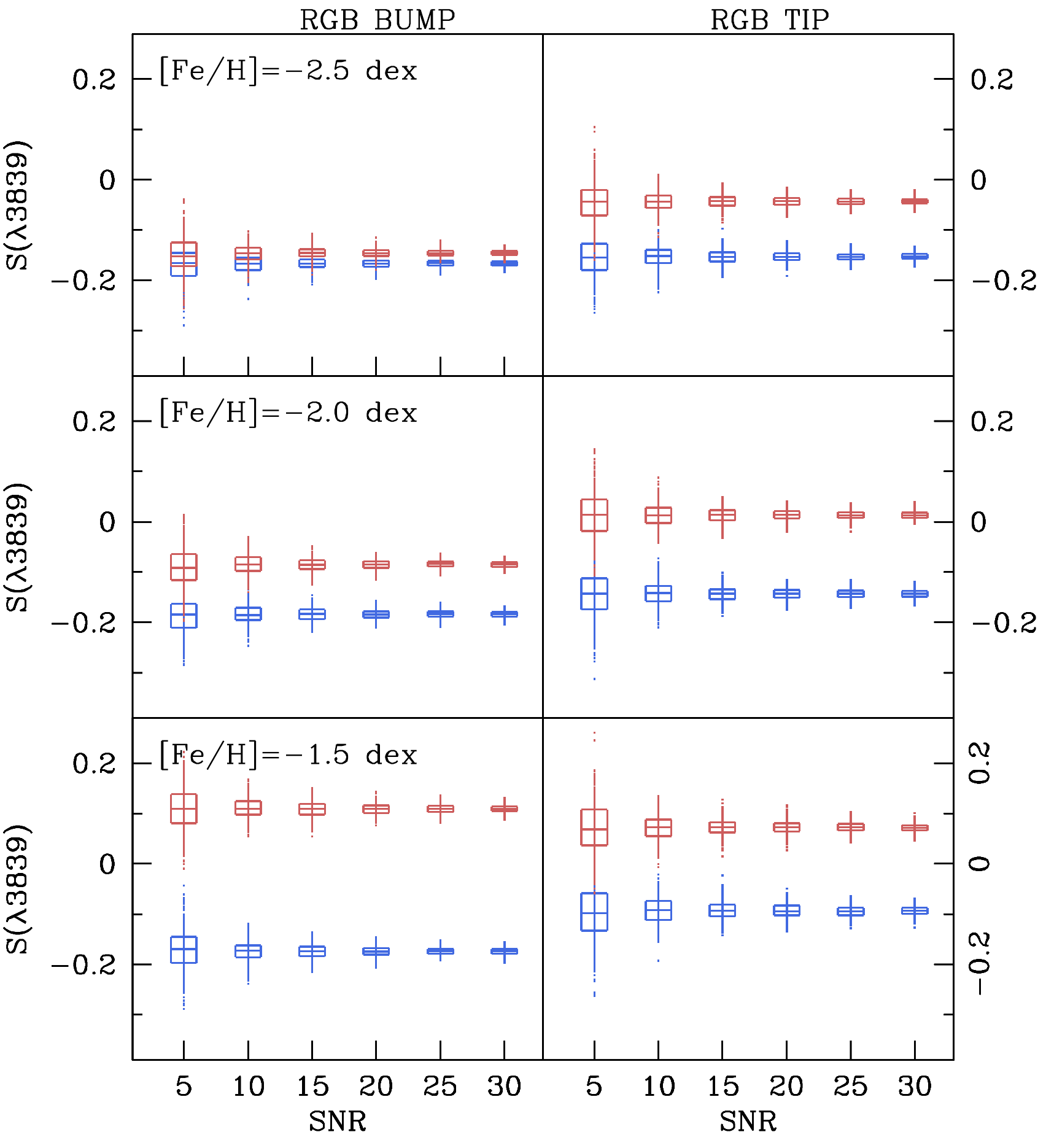}
\caption{Box and whisker plot of S($\lambda$3839) abundances 
as a function of the signal-to-noise ratio (per pixel) for 
a CN-normal (blue) and CN-strong (red) simulated stars. Left-hand panels are the results of template 
synthetic spectra with typical 
atmospheric parameters as for RGB bump stars, while right-hand panels indicate a representative star at the tip of the RGB.
For each evolutionary stage, we report results for three different metallicities, 
$[{\rm Fe/H}]=-2.5$, --2.0, and --1.5 (from top to bottom).
For a given SNR, the bottom and top of the box represents the 25\% and 75\% percentile range, respectively,  and the median is indicated by the band inside the box . The vertical tails extending
from the boxes indicate the total range of  S($\lambda$3839)  measurements determined for each SNR.}
\label{fig:NOISE}
\end{center}
\end{figure}

\begin{figure}
\begin{center}
\includegraphics[width=0.95\columnwidth]{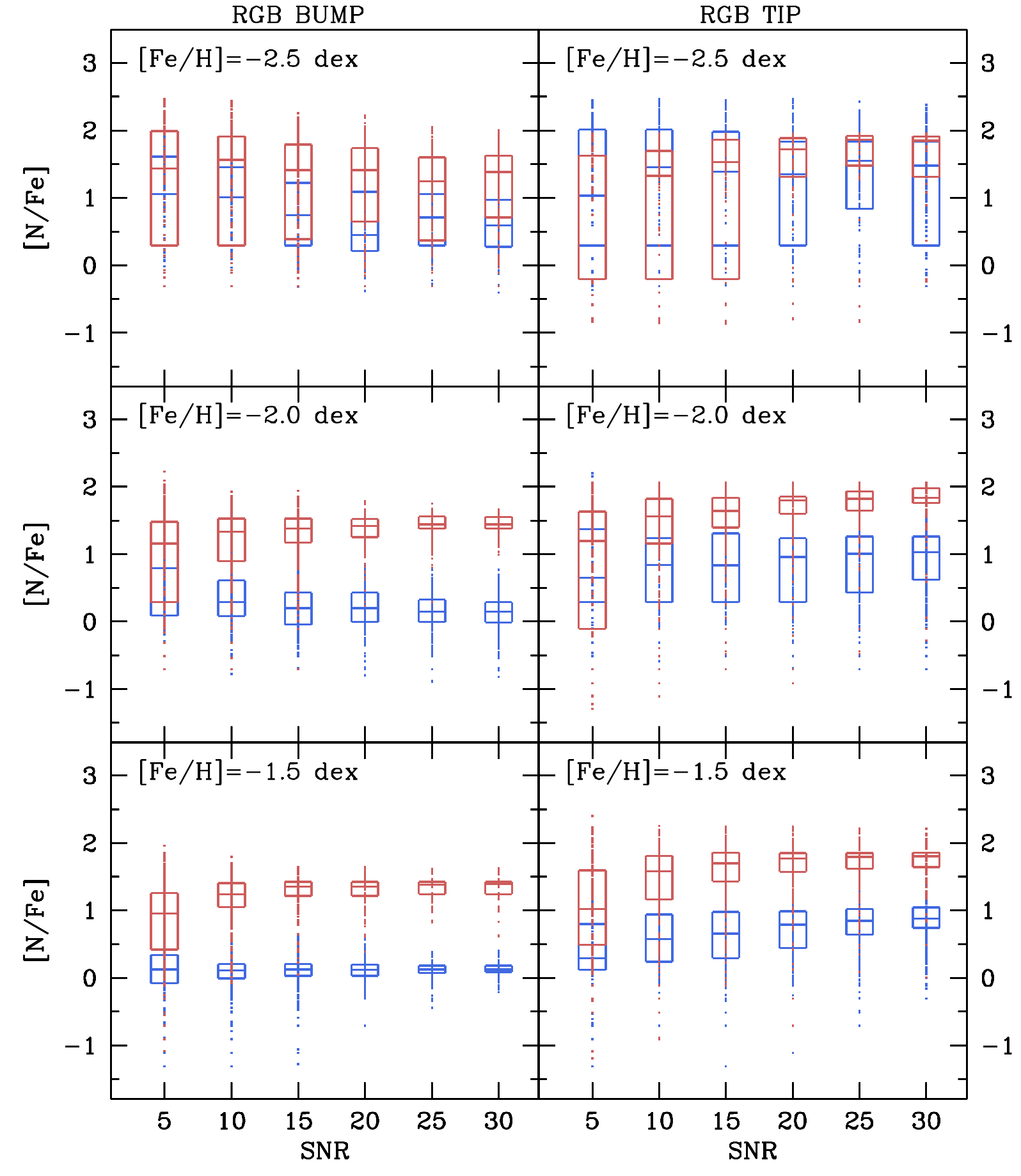}
\caption{Same as for Fig.~\ref{fig:NOISE}, except for  [N/Fe] abundances.}
\label{fig:ABUNOISE}
\end{center}
\end{figure}

We tested the hypothesis that the presence of CN-strong, N-rich stars at the low metallicity of the substructure 
could be hidden by  large measurement errors due to the low SNR of our spectra in the region of the CN band. 
In Figs.~\ref{fig:NOISE} and ~\ref{fig:ABUNOISE} we plot the S($\lambda$3839) band strength and the [N/Fe] abundance against the signal-to-noise ratio (per pixel) of two simulated CN-normal and CN-strong stars at a different evolutionary stage (at the RGB bump and at the tip of the RGB). 
For each SNR, we simulate 500 CN-normal and 500 CN-strong spectra with Poissonian noise. 
For bump stars, we assume [O/Fe]=+0.3, [C/Fe]=--0.4, [N/Fe]=+0.2 dex, and [O/Fe]=+0.4, [C/Fe]=--0.8, [N/Fe]=+1.4 dex for the 
CN-normal and CN-strong stars, respectively. Those chemical mixtures have been chosen to reproduce the typical ranges in 
C, N, and O abundances observed in unevolved GC stars at low-metallicity \citep{cohen05}. 
As a result of evolutionary effects during the RGB, we expect the  carbon abundance to decrease in value 
with increasing luminosity for stars located on the upper RGB  (e.g., \citealt{gratton00}).
Therefore for the representative tip star, we adopt  [O/Fe]=+0.4 dex, [C/Fe]=--1.0 dex,  [N/Fe]=+0.8 dex, and [O/Fe]=+0.4, [C/Fe]=--1.4, [N/Fe]=+1.8 dex for the CN-normal and CN-strong stars, respectively. 
In addition, we explore three metallicity regimes $[{\rm Fe/H}]=-2.5$, --2.0, and --1.5 dex (i.e., the metallicity range covered by our data). Finally, to compute synthetic spectra, we assume T$_{\rm{eff}}$= 5000K, $\log$(g)=2.0 dex, and $v_{\rm mic}$=1 kms$^{-1}$ for the bump star, while for a star near the tip we adopt T$_{\rm{eff}}$= 4000K, $\log$(g)=0.5 dex, and $v_{\rm mic}$=1 kms$^{-1}$.

As expected, from Figs.~\ref{fig:NOISE} and~\ref{fig:ABUNOISE}, we observe that even large variations in the [N/Fe] abundance (as large as $\Delta$[N/Fe]$\simeq$ 1.0-1.2 dex)  reflect a small separation between the CN-normal/N-poor and CN-strong/N-rich stars in the metal poor regime ([Fe/H] $\leq$ --2.0 dex). 
However, given the error bars, the separation between the CN-normal and CN-strong stars can already be seen 
at metallicities $\geq$--2.0 in the indexes of Fig.~\ref{fig:NOISE}; on the other hand, the larger measurement errors connected to the determination of 
N abundances from the spectral synthesis reflect into a less significant separation between the groups of N-poor and N-rich stars. 
{We conclude that the SNR of our data would have been sufficient to observe a (marginal) separation in the populations, at least in the S($\lambda$3839) index, at the very low metallicity characteristic of the chemodynamical substructure as well.}

In GCs the present-day typical fraction of CN-normal and CN-rich stars is $\sim$50\%, ranging from $\sim$30\% to 70\%
(e.g.,~\citealp{pancino10}; see also \citealp{bastian15} for a discussion about the fraction of enriched-to-normal stars). For a binomial distribution with a 50\% probability of success,
there is only a small, although not statistically negligible chance of not detecting any CN-strong stars 
(probability $\sim$0.016), which can increase (decrease) to 0.12 (0.0007) for  a  70\% (30\%) probability of success.
Given the uncertainties/intrinsic variations in the
fraction of CN-strong and CN-weak stars determined in GCs, we cannot exclude that our nondetection of
CN-strong stars may be due to small number statistics. 

Furthermore, various scenarios that explain chemical anomalies in GCs as the result of self-enrichment
postulate that GCs were originally from 10 to 100 times 
more massive than at present, where the missing mass would all be formed of first generation stars. These large mass
losses are in contrast with recent results on the Fornax dSph \citep{larsen12}, which set upper limits
of four to five times. Even assuming these large values, if the substructure we are seeing is a GC disrupted {before}
it could lose most of the first generation of stars, then the probability of detecting second generation
stars would have been even lower. It cannot also be excluded that we are only probing a limited fraction of the
stream of tidally shredded GC stars, which could all be made of first generation stars. 

Owing to small number statistics and limited spatial coverage as well as uncertainties in our understanding of
present-day GC properties and past chemical enrichment, we cannot exclude the possibility that we are
observing  a disrupted GC.

\section{Summary and conclusions}\label{SUMMARY}

We present abundances or upper limits for carbon and nitrogen for 94 RGB stars in Sculptor,
derived from low-resolution (R$\simeq$1150) spectroscopic observations with VLT/VIMOS. The
C and N abundances have been derived by fitting model spectra with appropriate 
atmospheric parameters to the G band at 4300~\AA~and CN band at 3883~\AA~, respectively.
Among these stars, only nine had previously determined carbon abundances.
To our knowledge, this is the first time that N abundances have been measured for a large number of giant stars in
a dSph galaxy. 
The main results of our analysis can be summarized as follows.

We measure the CH and CN band strengths and find no significant variations of the CH index strengths
with $V$ magnitude. On the contrary, we notice a trend of increasing CN index for brighter stars, with the relation being
steeper for increasing metallicities.

Carbon shows a decline with increasing luminosity in each metallicity bin.
Conversely, nitrogen increases with evolutionary stage along the RGB only in stars  
more metal-poor than [Fe/H]$\leq$--2.2 dex. We compare the measured abundances with theoretical models 
to investigate these trends and find excellent agreement between our [C/Fe] and [N/Fe] abundances and the \citet{placco14}
 theoretical predictions for the range of $\log$(g) values probed by our target stars.
 
The average value of [C/Fe], corrected for evolutionary effects, suggests that later forming stars 
(i.e., more metal-rich stars) started with lower [C/Fe] ratios than their predecessors.
Additionally, we detect a dispersion in C abundances that exceeds the scatter expected from measurement uncertainties alone.
This can be explained by the pollution from different nucleosynthesis sources over the time 
of interstellar matter, which additionally has not been well mixed. Although we only qualitatively discuss the 
evolution of  C abundances, the new data presented here should be used together with literature abundance determinations
\citep{kirby15}
to make quantitative predictions of the star formation history in Sculptor (following \citealp{romano13}).

We identify two new CEMP stars at [Fe/H]$\simeq$--2.2dex,
  with original carbon abundance ratios [C/Fe]$_{\rm{for}}$=1.49 and 1.38 dex.
  These stars were also identified as CN-strong stars with large CH absorptions. 
They are also enriched in Ba abundance [Ba/Fe]>1.0 dex, suggesting that both stars have been polluted
with a significant amount of carbon and $s$-process elements by an AGB companion. The updated statistics
of CEMP and CEMP-no stars in Sculptor are still compatible with a framework in which primordial faint supernovae
dominated the early chemical enrichment in the MW halo as well as in Sculptor \citep{salvadori15}.

Even if further observations are 
needed to determine elemental abundances of these stars in greater detail,
we demonstrated that low-resolution, multiobject spectroscopy of a large 
number of stars around the UV/CN band can be a quick way to identify candidate CEMP stars
for follow-up, high-resolution investigations.

None of the stars belonging to the kinematic substructure uncovered in Scl dSph by \citet{battaglia07}
  displays the chemical C-N signature typical of second populations stars born in GCs. 
  We demonstrated that, even at low metallicity of the substructure, the nondetection of CN-strong,
  N-rich stars cannot be attributed to 
insufficient S/N ratio if the variations in [C/Fe] and [N/Fe] in the {putative dissolved clusters}
are as large as those observed in metal-poor Galactic GCs 
(see \citealp{larsen14} for [N/Fe] variations in the GC system of Fornax).
However, because of the uncertainties in the properties and evolution of GCs, as well as
to the low number statistics and limited spatial coverage of our sample, we cannot exclude the GC hypothesis. 
Therefore, additional high-resolution studies designed to measure the key element Na are needed
to establish the nature of this feature.

%
%
%
%
%
%
%
%
%
%

\begin{acknowledgements} 
We are grateful to the anonymous referee for his/her careful reading of this manuscript.
CL and GB thank the INAF- Bologna Observatory for their hospitality during part of this
work. GB gratefully acknowledges support through a Marie-Curie action Intra European Fellowship, funded by the European
Union Seventh Framework Program (FP7/2007-2013)
under Grant agreement number PIEF-GA-2010-274151, as
well as the financial support from the Spanish Ministry of
Economy and Competitiveness (MINECO) under the Ramon
y Cajal Programme (RYC-2012-11537).
DR acknowledges financial support from PRIN MIUR 2010-2011, project ``The Chemical and Dynamical Evolution of the Milky Way and Local Group Galaxies'', prot. 2010LY5N2T.
The authors are indebted to the International Space Science Institute (ISSI), Bern, Switzerland, 
for supporting and funding the international team ``First stars in dwarf galaxies''. 
\end{acknowledgements}

\bibliographystyle{aa}
\bibliography{bibliography}

\end{document}